\documentclass{article}
\usepackage{bbm}
\usepackage{anysize}
\marginsize{3cm}{3cm}{4cm}{4cm}
\usepackage{makeidx}
\usepackage{amssymb}
\usepackage{amsmath}
\usepackage[Large]{subfigure}
\usepackage{graphicx}
\usepackage{lscape}
\usepackage{verbatim}
\usepackage{enumitem}
\usepackage[affil-it]{authblk}

\newtheorem{theorem}{Theorem}

\newtheorem{remark}[theorem]{Remark}

\makeindex
\begin{document}

\author{Stefano, M. Iacus%
  \thanks{Electronic address: \texttt{stefano.iacus@unimi.it}}}
\affil{Department of Economics, Management and Quantitative Methods\\
University of Milan\\ CREST Japan Science and Technology Agency\\}
\author{Lorenzo Mercuri 
\thanks{Electronic address: \texttt{lorenzo.mercuri@unimi.it}}}
\affil{Department of Economics Management and Quantitative Methods\\
University of Milan\\ CREST Japan Science and Technology Agency\\}
\author{Edit Rroji
\thanks{Electronic address: \texttt{e.rroji@unimib.it}}}
\affil{Department of Statistics and Quantitative Methods\\
 University of Milano-Bicocca\\}

\title{Estimation and Simulation of a COGARCH(p,q) model in the \texttt{YUIMA} project}
\maketitle
\abstract{In this paper we show how to simulate and estimate a COGARCH(p,q) model in the R package \texttt{yuima}. Several routines for simulation and estimation are available. Indeed for the generation of a COGARCH(p,q) trajectory, the user can choose between two alternative schemes. The first is based on the Euler discretization of the  stochastic differential equations that identifies a COGARCH(p,q) model while the second one considers the explicit solution of the variance process. \newline Estimation is based on the matching of the empirical with the theoretical autocorrelation function. In this case three different approaches are implemented: minimization of the mean square error, minimization of the absolute mean error and the generalized method of moments where the weighting matrix is continuously updated.\newline Numerical examples are given in order to explain methods and classes used in the \texttt{yuima} package.
}
\tableofcontents

%

\section{Introduction}
The Continuous-Time GARCH(1,1) process has been introduced in \cite{Cogarch2004} as a continuous counterpart of the discrete-time GARCH(1,1) model proposed by \cite{Bollerslev86}. \newline
The idea is to develop in continuous time a model that is able to capture some stylized facts observed in financial time series \cite{Cont01empiricalproperties} exploiting only one source of randomness for returns and for variance dynamics. Indeed, in the Continuous-Time GARCH (COGARCH hereafter), the stochastic differential equation for variance is driven by the discrete part of the quadratic variation of the same L\'evy process used for modeling returns. The continuous nature of the COGARCH makes it particularly appealing for discribing the behaviour of high frequency data [see \cite{Haug2007} for an application of method of moments using intraday returns].\newline
The generalization to higher order COGARCH(p,q) processes has been proposed in \cite{Brockwell2006,Chadraa2010Thesis}. Starting from the observation that the variance of a GARCH(p,q)   is an ARMA(q, p-1), the Variance is modeled with a CARMA(q,p-1) process  [see \cite{Brockwell2001,Tomasson2011,BrockwellDavisYang2007} and many others] driven by the discrete part of the quadratic variation of the L\'evy process in the returns. Although this representation is different from the one used by \cite{Cogarch2004} for the COGARCH(1,1) process, this last can be again retrieved as a special case. \newline
Many authors recently have investigated the COGARCH(1,1) model from a theoretical and an empirical point of view [see \cite{maller2008,Kallsen200974,muller2010mcmc,Bibbona2015} and many others]. Some R codes for estimation and simulation of a COGARCH(1,1) driven by a Compound Poisson and Variance Gamma are available in \cite{Granzer2013Thesis}. For the general COGARCH(p,q), the main contribution  remain the seminal works \cite{Brockwell2006} and \cite{Chadraa2010Thesis}. The aim of this paper is to describe the simulation and the estimation schemes in the \texttt{yuima} package \cite{yuimaPack} for a COGARCH(p,q) model driven by a general L\'evy process. Based on our knowledge \texttt{yuima} is the first R package available on CRAN that allows the user to manage a higher order COGARCH(p,q) model. Moreover, the estimation algorithm gives the option to recover the increments of the underlying noise process and estimates the L\'evy measure parameters. We recall that a similar procedure is available in \texttt{yuima} also for the CARMA model \cite[see][for a complete discussion]{IacusMercur2015}. The \texttt{yuima} package is developed within the \texttt{YUIMA} project \cite{Brousteetal2013} whose aim is to provide an environment for simulation and estimation of stochastic differential equations.\newline
The outline of the paper is as following. In Sect. \ref{TheoryCOGARCH} we discuss the main properties of the COGARCH(p,q) process. In particular we review the condition for existence of a strictly stationary variance process, its higher moments and the behaviour of the autocorrelation of the square increments of the COGARCH(p,q) model. In Sect. \ref{SimulatCogarch} we analyze two different simulation schemes. The first is based on the Euler discretization  while the second one uses the solution of the state process in the CARMA(q,p-1) model. Sect. \ref{TwoStepEstCOGARCH} is devoted to the estimation algorithm. In Sect. \ref{PackageRDescrip} we show the main classes and corresponding methods in \texttt{yuima} package and in the  Sect. \ref{NumExamp}
we present some numerical examples about the simulation and the estimation of a COGARCH(p,q) model

\section{COGARCH  Models driven by a L\'{e}vy process}
\label{TheoryCOGARCH}
In this section we review the mathematical definition of a COGARCH(p,q) process and its properties. In particular we focus on the conditions for the existence of a strictly stationary COGARCH(p,q) process and compute the first four unconditional moments. The existence of higher order moments plays a central role for the computation of the autocorrelation function of the squared increments of the COGARCH(p,q) model and consequently the estimation procedure implemented in the \texttt{yuima} package.\newline
The COGARCH(p,q) process, introduced in \cite{Brockwell2006} as a generalization of the COGARCH(1,1) model, is defined through the following system of stochastic differential equations:

\begin{equation}
\left\{\begin{array}{l} 
 \mbox{d}G_t =  \sqrt { V_{t} }  \mbox{d}L_t  \\ 
 V_t =  a_0 +   \mathbf{a}^{\intercal} Y_{t-} \\ 
 \mbox{d}Y_t = A Y_{t-}\mbox{d}t + \mathbf{e} \left(a_0 +   \mathbf{a}^{\intercal}Y_{t-}\right)\mbox{d}\left[ L,L \right]^{d}_t \\ 
 \end{array}\right.
 \label{def:Cogarch}
\end{equation}
where  $q$ and $p$ are integers such that $q \geq p \geq 1$. The state space process $Y_{t}$ is a vector with $q$ components:
\begin{equation*}
Y_t =\left[ Y_{1,t} ,  \ldots   ,  Y_{q, t} \right]^{\intercal}.
\end{equation*}
The vector $\mathbf{a} \in \mathcal{R}^{q}$ is defined as:
\begin{equation*}
\mathbf{a} = \left[ a_1 ,  \ldots   ,  a_{p}  ,a_{p+1}, \ldots , a_{q} \right]^{\intercal}
\end{equation*}
with $a_{p+1}=\dots=a_{q}=0$. The companion $q \times q$ matrix $A$  is
\begin{equation*}
A  = \left[ \begin{array}{cccc}
  0 & 1 & \ldots & 0 \\ 
 \vdots & \vdots & \ddots & \vdots \\ 
 0 & 0 & \ldots & 1 \\ 
 -b_q  &   -b_{q-1}  & \ldots &  -b_1 \\ 
 \end{array}\right].
\end{equation*}
The vector $\mathbf{e}\in \mathcal{R}^{q}$ contains zero entries except the last component that is equal to one. \newline
$\left[ L,L \right]^{d}_t$ is the discrete part of the quadratic variation of the underlying L\'evy process $L_{t}$ and is defined as:
\begin{equation}
\left[ L,L\right]^{d}_{t}:=\sum_{0\leq s \leq t}\left(\Delta L_{s}\right)^{2}.
\label{disc:QuadrVar}
\end{equation}
\begin{remark}
A COGARCH(p,q) model is constructed starting from the observation that in the GARCH(p,q) process, its discrete counterpart, the dynamics of the variance is a predictable ARMA(q, p-1) process driven by the  squares of the past increments. In the COGARCH(p,q) case, the ARMA process leaves the place to a CARMA(q,p-1) model [see \cite{Brockwell2001} for details about the CARMA(p,q) driven by a L\'evy process] and the infinitesimal increments of the COGARCH(p,q) are defined through the differential of the driven L\'evy  $L_{t}$ as done in \eqref{def:Cogarch}.   
\end{remark}
As observed above the COGARCH(p,q) model generalizes the COGARCH(1,1) process that has been introduced following different arguments from those for the $\left(p,q\right)$ case. However choosing $q=1$ and $p=1$ in \eqref{def:Cogarch} the COGARCH(1,1) process developed in \cite{Cogarch2004, Haug2007} can be retrieved through straightforward manipulations and, for obtaining the same parametrization in Proposition 3.2 of \cite{Cogarch2004}, the following equalities are necessary:
\begin{equation*}
\omega_{0} =a_{0}b_{1}, \ \omega_{1}=a_{1}e^{-b_{1}} \ \text{and} \ \eta=b_1.
\end{equation*}
Before introducing the conditions for strict stationarity and the existence of unconditional higher moments, it is worth noting that the state space process $Y_t$ can be seen as a Multivariate Stochastic Recurrence Equation and the corresponding theory can be applied to the COGARCH(p,q) process [see \cite{Brockwell2006,Chadraa2010Thesis} more details] in order to derive its main features\footnote{The Stochastic Recurrence Equations theory \cite{Brandt1986,Kesten1973} has been also used to prove the strictly and weakly stationarity for the GARCH(p,q) model \cite{Basrak2002} }. In the case of the Compound Poisson driven noise,  the representation through the stochastic difference equations is direct in the sense that the random coefficients of the state  process $Y_t$ can be written explicitly while in the general case, it is always possible to identify a sequence of Compound Poisson processes that as limit to the choosen driven L\'evy process.

In the following, we require that matrix $A$ can be diagonalized, that means:
\begin{equation*}
A= SDS^{-1}
\end{equation*}
with
\begin{equation}
S=\left[ 
\begin{array}{ccc}
1 & \ldots & 1\\
\lambda_{1} & \ldots & \lambda_q\\
\vdots & & \vdots\\
\lambda^{q-1}_{1} & \ldots & \lambda^{q-1}_{q}\\
\end{array}
\right], \ D=\left[\begin{array}{ccc}
\lambda_{1}& &\\
& \ddots &\\
& & \lambda_{q}\\
\end{array}\right]
\label{DiagMatrixCond}
\end{equation}
where the $\lambda_{1},\lambda_{q},\ldots,\lambda_q$ are the eigenvalues of matrix $A$ and are ordered as follows:
\begin{equation*}
\Re\left\{\lambda_{1}\right\}\geq \Re\left\{\lambda_2\right\}\geq \ldots \geq \Re\left\{\lambda_q\right\}.
\end{equation*}
Applying the theory of stochastic recurrence equations, \cite{Brockwell2006} provide A sufficient condition for the strict stationarity of  a COGARCH(p,q) model. We review the result for the stationarity of the state process $Y_t$. Two fundamental assumptions are the fact that the eigenvalues $\lambda_{1},\ldots,\lambda_{q}$ are distinct and the underlying process $L$ must have a non-trivial $\nu_{L}\left(l\right)$ measure. Then, the  process $Y_t$ converges in distribution to the random variable $Y_{\infty}$   if exist some $r\in\left[ 1,+\infty \right]$ such that:
\begin{equation}
\int_{-\infty}^{+\infty}\ln\left(1+\lVert S^{-1}\mathbf{e}\mathbf{a}^{\intercal}S\lVert_{r}l^2\right)\mbox{d}\nu_{L}\left(l\right) \leq \Re\left\{\lambda_{1}\right\}
\label{statSol}
\end{equation}
for some matrix $S$ such that the matrix $A$ is diagonalizable. If we choose as a starting condition $Y_0\overset{d}=Y_{\infty}$ than the process $Y_t$ is strictly stationary and consequently the variance $V_t$ is also a strictly stationary process. Unlikely for the general case, the inequality in \eqref{statSol} gives only a sufficient condition about the strict stationarity, and it is difficult to verify in practice\footnote{In the \texttt{yuima} package a diagnostic for condition \eqref{statSol} is available choosing matrix $S$ as done in \eqref{DiagMatrixCond} and $r=2$. We remark that the process $Y_t$ is stationary if the diagnostic gives a positive answer otherwise we can conclude nothing about the stationarity of the process.}. As shown in \cite{Cogarch2004} and remarked in \cite{Brockwell2006},  the condition in \eqref{statSol} is also necessary for the COGARCH(1,1) case and  can be simplified as:
\begin{equation}
\int_{-\infty}^{+\infty}\ln\left(1+a_{1}l^2\right) \mbox{d}\nu_{L}\left(l\right) \leq b_1.
\label{cogarch11StatSol}
\end{equation}
Using again the SRE theory, it is also possible to determine the condition for existence of higher moments of the state process $Y_t$. In this way it is possible to determine the autocorrelations of the squared COGARCH increments that are used in the estimation procedure illustrated in Section \ref{TwoStepEstCOGARCH}. As reported in \cite{Brockwell2006}, the  $\kappa$-th order moment of the process $Y_t$ exists finite if, for some $r\geq 1$ and $S$ such that the matrix $A$ is diagonalizable, the following conditions hold:
\begin{equation}
E\left(L_1^{2\kappa}\right)<+\infty, \ \int_{-\infty}^{+\infty}\left[\left(1+\lVert S^{-1}\mathbf{e}\mathbf{a}^{\intercal}S \rVert_{r}l^{2}\right)^{\kappa}-1\right]\mbox{d}\nu_L\left(l\right)<\Re\left\{\lambda_{1}\right\} \kappa.
\label{existkappaMoM}
\end{equation}
As special case of \eqref{existkappaMoM}, the unconditional stationary mean $E\left(Y_{\infty}\right)=-a_0 \mu \left(A+\mu \mathbf{e}\mathbf{a}^{\intercal}\right)^{-1}\mathbf{e}$ of the vector process $Y_t$ exists if 
\begin{equation*}
E\left(L^2_1\right)<+\infty,\ \lVert S^{-1} \mathbf{e}\mathbf{a}^{\intercal}S \rVert_{r}\mu < \Re\left\{ \lambda_{1}\right\}
\label{Cond2MoM}
\end{equation*}
where 
\begin{equation}
\mu:=\int_{-\infty}^{+\infty}l^2\mbox{d}\nu_{L}\left(l\right)
\end{equation}
is the second moment of the L\'evy measure $\nu_{L}\left(l\right)$.
It is worth noting that the condition \eqref{Cond2MoM} ensures also the strict stationarity since, using $\ln(1+x) \leq x $,  we have the following sequences of inequalities:
\begin{equation*}
\int_{-\infty}^{+\infty}\ln\left(1+\lVert S^{-1}\mathbf{e}\mathbf{a}^{\intercal}S\lVert_{r}l^2\right)\mbox{d}\nu_{L}\left(l\right) \leq \lVert S^{-1} \mathbf{e}\mathbf{a}^{\intercal}S \rVert_{r}\mu \leq \Re \left\{\lambda_{1}\right\}.
\end{equation*}
For the stationary covariance matrix \cite{Chadraa2010Thesis}
\begin{equation}
cov\left(Y_{\infty}\right)=\frac{a_{0}^2 b^{2}_q \rho}{\left(b_q-\mu a_1\right)^2\left(1-m\right)}\int_{0}^{\infty}e^{\tilde{A}t}\mathbf{e}\mathbf{e}^{\intercal}e^{\tilde{A}^{\intercal}t}\mbox{d}t,
\end{equation} 
the existence of the second moment of $Y_{\infty}$ in \eqref{existkappaMoM} becomes:
\begin{equation*}
E\left(L^4_1\right)<\infty,\ \lVert S^{-1} \mathbf{e}\mathbf{a}^{\intercal}S \rVert_{r} \rho < 2\left(-\Re\left\{\lambda_{1}\right\} - \lVert S^{-1} \mathbf{e}\mathbf{a}^{\intercal}S \rVert_{r}\mu\right)
\end{equation*}
where
\begin{equation*}
\rho:=\int_{-\infty}^{+\infty}l^4\mbox{d}\nu_{L}\left(l\right)
\end{equation*}
is the fourth moment of the L\'evy measure $\nu_{L}\left(l\right)$ and 
\begin{equation*}
m:=\int^{+\infty}_{0}\mathbf{a}^{\intercal}e^{\tilde{A}t}\mathbf{e}\mathbf{e}^{\intercal}e^{\tilde{A}^{\intercal}t}\mathbf{a}\mbox{d}t.
\end{equation*}
Before introducing the higher moments and the autocorrelations, we recall the conditions for the nonegativity for a strictly stationary variance process in \eqref{def:Cogarch}. Indeed, under the assumption that all the eigenvalues of matrix $A$ are distinct and the relation in \eqref{statSol} holds, the variance process $V_t \geq a_0>0$ a.s. if:
\begin{equation}
\mathbf{a}^{\intercal}e^{At}\mathbf{e}\geq 0, \ \forall t \geq 0.
\label{eq:nonNegVar}
\end{equation}
The condition in \eqref{eq:nonNegVar} is costly since we need to check it each time. Nevertheless some useful controls are available \cite{Tsai2005}.
\begin{enumerate}
\item A necessary and sufficient condition to guarantee that  $a^{\intercal}e^{At}\mathbf{e}\geq 0$
  in the COGARCH(2,2) case is that the eigenvalues of $A$ are real and $a_{2}\geq 0$ and $a_{1}\geq-a_{2}\lambda\left(A\right)$ where $\lambda\left(A\right)$ is the biggest eigenvalue.
\item Under condition $2\leq p\leq q$, that all eigenvalues of $A$ are negative and ordered in an increasing way $\lambda_{1}\geq\lambda_{2}\geq,\ldots,\geq\lambda_{p-1}$ and $\gamma_{j}$ are the roots of $a\left(z\right)=0$ ordered as $0>\gamma_{1}\geq\gamma_{2}\geq\ldots\geq\gamma_{p-1}$. Then sufficient condition for \eqref{eq:nonNegVar} is 
\begin{equation*}
  \sum_{i=1}^{k}\gamma_{i}\leq\sum_{i=1}^{k}\lambda_{i}\ \ \forall k\in\left\{1,\ldots,p-1\right\} .
\end{equation*}
\item For a COGARCH(1,q) model a sufficient condition that ensures \eqref{eq:nonNegVar} is that all eigenvalues must be real and negative.
\end{enumerate}
Combining the requirement in \eqref{statSol} with that in \eqref{eq:nonNegVar} it is possible to derive the higher moments and the autocorrelations for a COGARCH(p,q) model. As a first step, we define the returns of a COGARCH(p,q) process on the interval $\left(t,t+r\right], \ \forall t\geq 0$ as: 
\begin{equation}
G^{\left(r\right)}_{t}:= \int_{t}^{t+r} \sqrt{V_{s}}\mbox{d}L_{s}.
\label{eq:G_r}
\end{equation}
Let $L_t$  be a symmetric and centered L\'evy process such that the fourth moment of the associated L\'evy measure is finite,  we define the matrix $\tilde{A}$ as:
\begin{equation*}
\tilde{A}:=A+\mu \mathbf{e} \mathbf{a}^{\intercal} .
\end{equation*}
It is worth noting that the structure of $\tilde{A}$ is the same of $A$ except for the last row $q$ where $\tilde{A}_{q,}=\left(-b_q+\mu a_{1},\ldots, -b_1+\mu a_{q}\right)$. For any $t \geq 0$ and for any $h \geq r >0 $, the first two moments of \eqref{eq:G_r} are 
\begin{equation}
E\left[\left(G^{(r)}_{t}\right)\right]=0
\end{equation}
and
\begin{equation}
E\left[\left(G^{(r)}_{t}\right)^{2}\right]=\frac{\alpha_{0}b_{q}r}{b_{q}-\mu a_{1}}E\left[L_{1}^{2}\right].
\label{secondMom}
\end{equation}
The computation of the autocovariances and variance for squared returns \eqref{eq:G_r} require the following quantities defined as:
\begin{equation}
\begin{array}{l}
P_0 = 2\mu^2\left[3 \tilde{A}^{-1}\left( \tilde{A}^{-1} \left(e^{\tilde{A}r} I\right)-rI\right)-I\right]cov\left(\epsilon_{\infty}\right)\\
P_h = \mu^2 e^{\tilde{A}h} \tilde{A}^{-1}\left(I-e^{\tilde{A}r}\right)\tilde{A}^{-1}\left(e^{\tilde{A}r}-I\right)cov\left(\epsilon_{\infty}\right)\\
\end{array},
\end{equation}
\begin{equation}
\begin{array}{l}
Q_0= 6\mu\left[\left(rI-\tilde{A}^{-1} \left(e^{\tilde{A}r}-I\right)\right)cov\left(\epsilon_{\infty}\right) - \tilde{A}^{-1}\left(\tilde{A}^{-1} \left(e^{\tilde{A}r}-I \right)-rI\right)cov\left(\epsilon_{\infty}\right)\tilde{A}^{\intercal} \right]\mathbf{e} \\
Q_h=\mu e^{\tilde{A}h} \tilde{A}^{-1}\left(I-e^{-\tilde{A}r}\right) \left[\left(I-e^{\tilde{A}r}\right)-\tilde{A}^{-1}\left(e^{\tilde{A}r}-I\right)cov\left(\epsilon_{\infty}\right)\tilde{A}^{\intercal}\right]\mathbf{e}
\end{array}
\end{equation}
and
\begin{equation}
R=2r^2\mu^2+\rho.
\end{equation}
The terms $P_h$ and $P_0$ are $q \times q $ matrices. $Q_h$ and $Q_0$ are $q \times 1$ vectors and the term $R$ is a scalar. The $q \times q$ matrix $cov\left(\epsilon_{\infty}\right)$ is defined as: 
\begin{equation}
cov\left(\epsilon_{\infty}\right)=\rho \int_{0}^{\infty}e^{\tilde{A}t}\mathbf{e}\mathbf{e}^{\intercal}e^{\tilde{A}^{\intercal}t}\mbox{d}t
\end{equation}
where $\mu, \rho$ and $m$ have been defined before. \newline
Using the $q \times q$ matrix $P_h$ and the $q \times 1$ vector $Q_h$ the autocovariances of the squared returns \eqref{eq:G_r} are defined as:
\begin{equation}
\gamma_{r}\left(h\right):=cov\left[(G^{(r)}_{t})^2,(G^{(r)}_{t+h})^2\right]=\frac{a_0^2b_q^2\left(\mathbf{a}^{\intercal}P_h \mathbf{a}+\mathbf{a}Q_h\right)}{\left(1-m\right)\left(b_q-\mu a_1\right)^2}, \ \ h \geq r
\label{autocov:G}
\end{equation}
while the variance of the process $\left(G^{\left(r\right)}_t\right)^{2}$ is
\begin{equation}
\gamma_{r}\left(0\right):= var\left(\left(G^{\left(r\right)}_t\right)^{2}\right)=\frac{a_0^2b_q^2\left(\mathbf{a}^{\intercal}P_0 \mathbf{a}+\mathbf{a}Q_0+ R\right)}{\left(1-m\right)\left(b_q-\mu a_1\right)^2}.
\label{VARIANCEMoM}
\end{equation}
Combining the autocovariances in \eqref{autocov:G} with \eqref{secondMom} and \eqref{VARIANCEMoM} we obtain the autocorrelations:
\begin{equation}
\rho_{r}\left(h\right):=\frac{\gamma_{r}\left(h\right)}{\gamma_{r}\left(0\right)}=\frac{\left(\mathbf{a}^{\intercal}P_h \mathbf{a}+\mathbf{a}Q_h\right)}{\left(\mathbf{a}^{\intercal}P_0 \mathbf{a}+\mathbf{a}Q_0+ R\right)}.
\label{COGARCH:autocorr}
\end{equation}
We conclude this Section with the following remark. As done for the GARCH(p,q) model the autocovariances of the returns $G^{\left(r\right)}_t$ are all zeros.

\section{Simulation of a COGARCH(P,Q) model}
\label{SimulatCogarch}
In this Section we illustrate the theory behind the simulation routines available in the \texttt{yuima} package when a COGARCH(p,q) model is considered. The corresponding sample paths are generated according two different schemes.\newline The first method, in the \texttt{Yuima} project \cite{Brousteetal2013}, is based on the Euler-Maruyama \cite{R:Iacus:2007} discretization of the system in \eqref{def:Cogarch}. In this case the algorithm follows these steps:
\begin{enumerate}
\item We determine an equally spaced grid from $0$ to $T$ composed by $N$ intervals with the same length  $\Delta t$
\begin{equation*}
0,\ldots,\left(n-1\right) \times \Delta t, n \times \Delta t, \ldots, N \times \Delta t. 
\end{equation*}
\item We generate the trajectory of the underlying L\'evy process $L_{t}$ sampled on the grid defined in step 1.
\item  Choosing a starting point for the state process $Y_0=y_0$ and $G_0=0$, we have 
\begin{equation}
Y_{n}=\left(I+A\Delta t\right)Y_{n-1}+\mathbf{e}\left(a_{0}+\mathbf{a}^{\intercal}Y_{n-1}\right)\Delta \left[LL\right]_{n}^{d}.
\label{condStatDiscr}
\end{equation}
The discrete quadratic variation of the $L_t$ process is approximated by
\begin{equation}
\Delta \left[LL\right]_{n}^{d}=\left(L_{n}-L_{n-1}\right)^2.
\end{equation}
\item Once the approximated state process $Y_n$ is obtained we can generate the trajectory of Variance and the process $G_n$ according the following equations:
\begin{equation}
V_n=a_0+\mathbf{a}^{\intercal} Y_{n-1}
\end{equation}
and
\begin{equation}
G_{n}=G_{n-1}+\sqrt{V_n}\left(L_n-L_{n-1}\right).
\end{equation}
\end{enumerate}

It is worth noting that, although the discretized version of the state process $Y_n$ in \eqref{condStatDiscr} can be seen as a stochastic recurrence equation, the conditions for stationarity and non-negativity for the variance $V_n$ process are not the same of ones analyzed in the Section \ref{TheoryCOGARCH}. In particular it is possible to determine an example where the discretized variance process $V_n$ assumes negative values while the true process is non negative with probability one.\newline In order to clarify deeper this issue we consider a COGARCH(1,1) model driven by a Variance Gamma L\'evy process [see \cite{Madan1990,Loregian2012} for more details about the VG model]. In this case, the condition for the non-negativity of the Variance in \eqref{eq:nonNegVar} is ensured if $a_{0}>0$ and $a_{1}>0$ while the strict stationarity condition in \eqref{cogarch11StatSol} for the COGARCH(1,1) is  $E[L^2]=1$ and $a_1-b_1<0$. The last two requirements guarantee also the existence of the stationary unconditional mean for the  process $V_{t}$. We define the model using the \texttt{yuima} function \texttt{setCogarch}. Its usage is completely explained in the Section \ref{PackageRDescrip}. The following command line instructs \texttt{yuima} to build the COGARCH(1,1) model with VG noise:
\begin{verbatim}
> model1 <- setCogarch(p = 1, q = 1, work = FALSE,
+         measure=list("rngamma(z, 1, sqrt(2), 0, 0)"), measure.type = "code",
+         Cogarch.var = "G", V.var = "v", Latent.var="x", XinExpr=TRUE)
\end{verbatim}
Choosing the following values for the parameters
\begin{verbatim}
> param1 <- list(a1 = 0.038, b1 = 301, a0 =0.01, x01 = 0)
\end{verbatim}
the COGARCH(1,1) is stationary and the variance is strictly positive. Nevertheless, if we simulate the trajectory using the Euler discretization, the value of $\Delta t$ can lead to negative values for the process as shown in the example:
\begin{verbatim}
> Terminal1=5
> n1=750
> samp1 <- setSampling(Terminal=Terminal1, n=n1)
> set.seed(123)
> sim1 <- simulate(model1, sampling = samp1, true.parameter = param1,
+         method="euler")
> plot(sim1, main="Sample Path of a VG COGARCH(1,1) model with Euler scheme")
\end{verbatim}

\begin{figure}[h]
	\centering
		\includegraphics[width=0.50\textwidth]{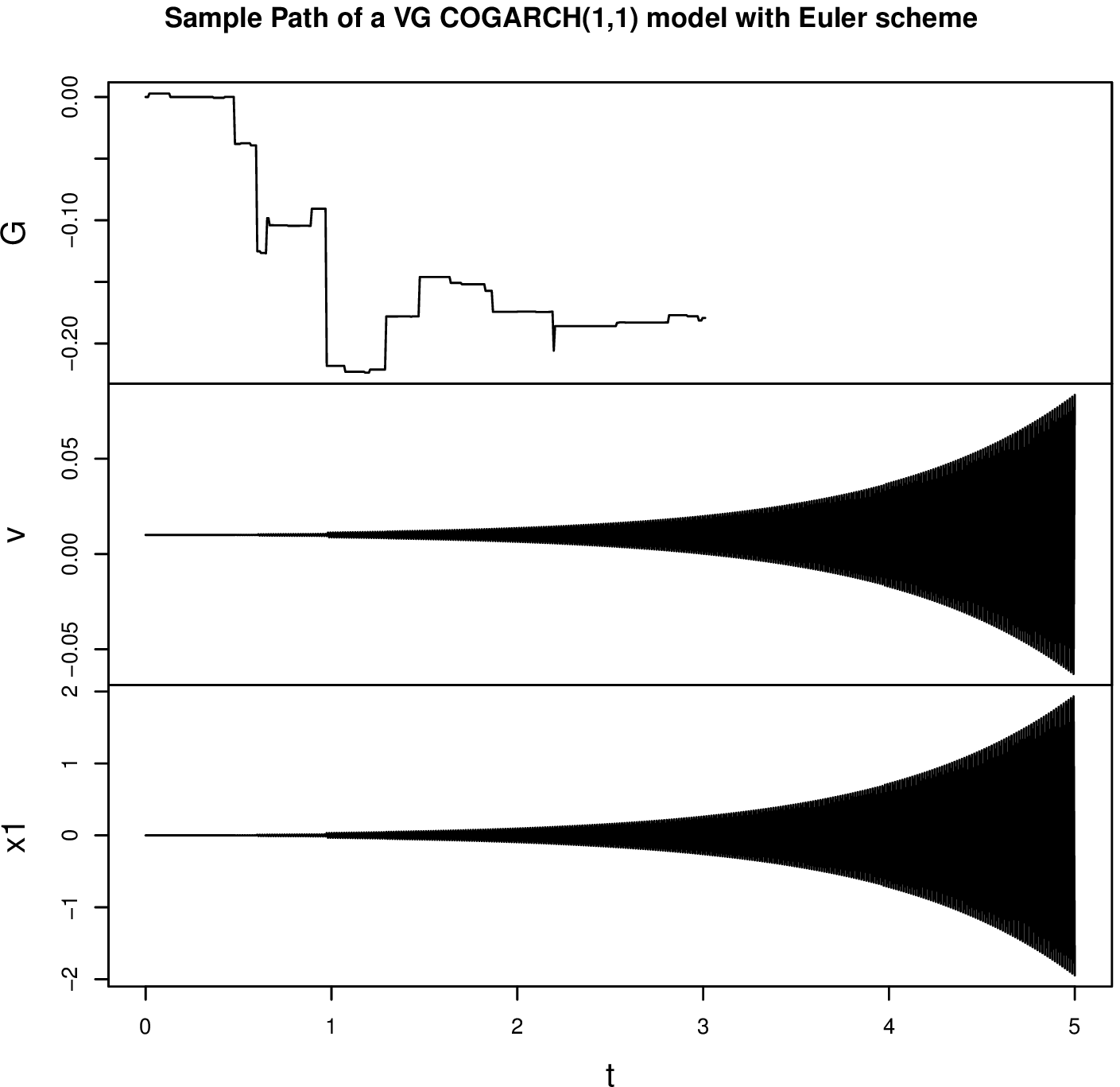}
\end{figure}

Looking to the Figure, we observe a divergent and oscillatory behaviour for the simulated  state $Y_n$ and the variance $V_n$ processes while, from theoretical point of view, the conditions for nonnegativity and stationarity for the variance of a COGARCH(1,1) model are satisfied by the values used for the parameters. \newline To overcome this problem, we provide an alternative simulation scheme based on the solution of the process $Y_n$ given the starting point $Y_{n-1}$.

Applying the Ito's Lemma for semimartingales \cite{protter1990stochastic} to the transformation $e^{-At}Y_t$, we have: 
\begin{equation*}
e^{-A \Delta t}Y_{t}=Y_{t- \Delta t}-\int_{t-\Delta t}^{t}Ae^{-Au}Y_{u-}\mbox{d}u+\int_{t- \Delta t}^{t}e^{-Au}\mbox{d}Y_{u}+\sum_{s\leq t}\left[e^{-As}\left(Y_{s}-Y_{s-}\right)-e^{-As}\left(Y_{s}-Y_{s-}\right)\right].
\end{equation*}
We substitute the definition of $Y_t$ in \eqref{def:Cogarch} and get 
\begin{equation*}
e^{-At}Y_{t}=Y_{t- \Delta t}-\int_{t- \Delta t}^{t}Ae^{-Au}Y_{u-}\mbox{d}u+\int_{t- \Delta t}^{t}e^{-Au}AY_{u-}\mbox{d}u+\int_{t- \Delta t}^{t}e^{-Au}e\left(a_{0}+a'Y_{u-}\right)\mbox{d}\left[LL\right]_{u}^{d}
\end{equation*}
Using the following property for an exponential matrix 
\begin{eqnarray*}
Ae^{At}&=&A\left(I+At+\frac{1}{2}A^{2}t^{2}+\frac{1}{3!}A^{3}t^{3}+...\right)\\
&=&\left(A+tA^{2}+\frac{1}{2}t^{2}A^{3}+t^{3}\frac{1}{3!}A^{4}+...\right)\\
&=&\left(I+tA+\frac{1}{2}t^{2}A^{2}+t^{3}\frac{1}{3!}A^{3}+...\right)A=e^{At}A,
\end{eqnarray*}
we get
\begin{equation}
Y_{t}=e^{At}Y_{t- \Delta t}+\int_{t- \Delta t}^{t}e^{A\left(t-u\right)}e\left(a_{0}+a'Y_{u-}\right)\mbox{d}\left[LL\right]_{u}^{d}
\label{explicitSolu}
\end{equation}
Except for the case where the noise is a Compound Poisson, the simulation scheme follows the same steps of the Euler-Maruyama discretization  where the state space process $Y_n$ on the sample grid is generated according the approximation of the relation in \eqref{explicitSolu}:
\begin{equation}
Y_{n} = e^{A\Delta t}Y_{n-1}+e^{A\left(\Delta t\right)}e\left(a_{0}+a'Y_{n-1}\right)\left(\left[LL\right]_{n}^{d}-\left[LL\right]_{n-1}^{d}\right)
\label{apprMixed0}
\end{equation} 
or equivalently:
\begin{equation}
Y_{n} = a_{0}e^{A\left(\Delta t\right)}e\Delta\left[LL\right]_{n}^{d}+e^{A\Delta t}\left(I+ea'\Delta\left[LL\right]_{n}^{d}\right)Y_{n-1}.
\label{apprMixed}
\end{equation} 
where $\Delta\left[LL\right]^{\left(d\right)}_{n}:= \left[LL\right]^{d}_{n}-\left[LL\right]^{d}_{n-1}$ is the increment of the discrete part of quadratic variation.\newline 
In the previous example, the sample path is simulated according the recursion in \eqref{apprMixed} choosing $\texttt{method = mixed}$ in the \texttt{simulate} function as done below:
\begin{verbatim}
> set.seed(123)
> sim2 <- simulate(model1, sampling = samp1, true.parameter = param1,
+         method="mixed")
> plot(sim2, main="Sample Path of a VG COGARCH(1,1) model with mixed scheme")
\end{verbatim}

\begin{figure}[h]
	\centering
		\includegraphics[width=0.50\textwidth]{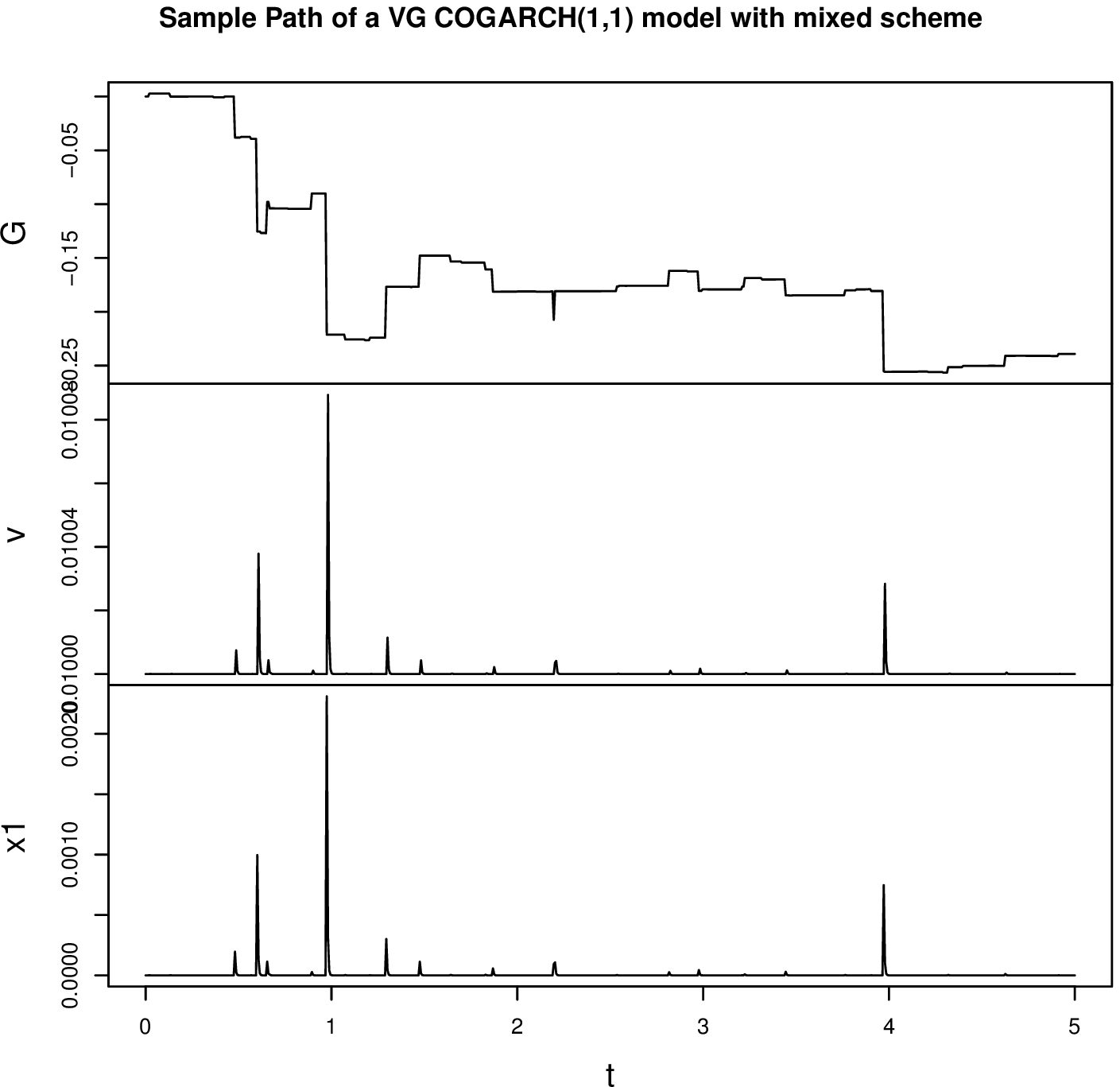}
\end{figure}
%

In the case of the COGARCH(p,q) driven by a Compound Poisson L\'evy process, a trajectory can be simulated without any approximation of the solution in \eqref{explicitSolu}. Indeed it is possible to determine the time where the jumps occur\footnote{In a general Compound Poisson the jump time follow an exponential r.v. with rate $\lambda$} and then evaluate the corresponding quadratic variation in an exact way. Once the trajectory of a random time is obtained, the piecewise constant interpolation is used on the fixed grid, in order to mantain the c\`adl\`ag property of the sample path.

\section{Estimation of a COGARCH(P,Q) model in the \texttt{yuima} package}
\label{TwoStepEstCOGARCH}
In this Section we explain the estimation procedure that we propose in the \texttt{yuima} package for the COGARCH(p,q) model. As done for the CARMA(p,q) model driven by L\'evy process \cite{IacusMercur2015}, even in this case a three step estimation procedure is proposed that allows the user to obtain estimated values for the COGARCH(p,q) and the parameters of the L\'evy measure. \newline This procedure is structured as follows:
\begin{enumerate}
\item Using the moment matching procedure explained below, we estimate the COGARCH(p,q) parameters $\mathbf{a}:=\left[ a_{1},\ldots, a_{p}\right]$, $\mathbf{b}:=\left[ b_{1},\ldots, b_{q}\right]$ and the constant term $a_0$ in the Variance process $V_t$. In this phase, the estimation is obtained by minimizing some distances between the empirical and theoretical autocorrelation function.  
\item Once the COGARCH(p,q) parameters are available, we recover the increments of the underlying L\'evy process using the methodology describe below.
\item In the third step, we use the increments obtained in the second step and estimate the L\'evy measure parameters by means of maximum likelihood estimation procedure.
\end{enumerate}

Let $G_{0},G_{1},\ldots,G_{n},\ldots, G_{T}$ be the observed values of the process subsampled at equally spaced instants $0,h,2h,\ldots,Nh$ where the length of each interval is $h=\frac{T}{N}$. The time of the last observation is $T$ and $N$ is the number of observations of the process $G_{n}$. \newline In the following we assume that the underlying L\'evy process is symmetric and centered in zero.
Starting from the sample $\left\{G_{n}\right\}^{N}_{n=0}$ we define the COGARCH(p,q) increments of lag one as:
\begin{equation*}
G^{\left(1\right)}_{n}:= G_{n}-G_{n-1}
\label{eq:deltaGlag1}
\end{equation*}
and the increments of lag $r$ as:
\begin{equation}
G^{\left(r\right)}_{n}:= G_{n}-G_{n-r} 
\label{eq:deltaGlagr}
\end{equation}
where $r\geq 1$ is a natural number and for $r=1$ the definition in \eqref{eq:deltaGlag1} coincides with \eqref{eq:deltaGlagr}.\newline  
It is worth mentioning that the increments $G^{\left(r\right)}_{n}$ can be obtained as a time aggregation of increments of lag one as follows:
\begin{equation}
G^{\left(r\right)}_{n} = \sum_{h=n-r}^{n} G^{\left(1\right)}_{n}
\label{eq:aggr}
\end{equation}
The time aggregation in \eqref{eq:aggr} can be useful during the optimization routine when the values of increments $ G^{\left(1\right)}$ are very small in absolute value. \newline
Using the sample $\left\{G^{\left(r\right)}_{n}\right\}_{n\geq r}$ we compute 
the empirical second moment 
\begin{equation*}
\hat{\mu}_{r}:=\frac{1}{N-d-r+1}\sum^{N-d}_{n = r} \left(G^{\left(r\right)}_{n}\right)^{2}
\end{equation*}
and empirical autocovariance function $\hat{\gamma}\left(h\right)$ is defined as:
\begin{equation*}
\hat{\gamma}_{r}\left(h\right):=\frac{1}{N-d-r+1}\sum^{N-d}_{n = r}\left(\left(G^{\left(r\right)}_{n+h}\right)^{2}-\hat{\mu}_{r}\right)\left(\left(G^{r}_{n}\right)^2-\hat{\mu}_{r}\right) \ \ h=0,1,\ldots,d
\end{equation*}
where $d$ is the maximum lag considered.\newline
The empirical autocorrelations are:
\begin{equation}
\hat{\rho}_{r}\left(h\right)=\frac{\hat{\gamma}_{r}\left(h\right)}{\hat{\gamma}_{r}\left(0\right)}.
\label{eq:empcorr}
\end{equation}
We use the theoretical and empirical autocorrelations in order to define the moment conditions for the estimation procedure. By the introduction of the $q+p$ vector $\theta:=\left(\mathbf{a},\mathbf{b}\right)$ we define the vector function $g\left(\theta\right): \mathbb{R}^{q+p} \rightarrow \mathbb{R}^d$ as follows:
\begin{equation}
g\left(\theta\right):=E\left[ f\left(G^{\left(r\right)}_{n},\theta\right)\right]
\label{eq:MoMconds}
\end{equation}
where  $f\left(G^{\left(r\right)}_{n},\mathbf{a},\mathbf{b}\right)$ is a $d$ dimensional real function where the components are defined as:
\begin{equation}
f_{h}\left(\Delta G^{\left(r\right)}_{n},\theta\right)=\rho_{r}\left(h\right)-\frac{\left( \left(G^{\left(r\right)}_{n+h}\right)^2-\mu_{r}\right)\left( \left(G^{\left(r\right)}_{n}\right)^2-\mu_{r}\right)}{\gamma_{r}\left(0\right)}, \ \ h=1, \ldots, d.
\end{equation}
In the estimation algorithm, we replace the expectation in \eqref{eq:MoMconds} with the sample counterpart. The components of vector $\hat{g}\left(\theta\right)=\left[ \hat{g}_{1}\left(\theta\right),\ldots,\hat{g}_{d}\left(\theta\right) \right]^{\intercal}$  are defined as: 
\begin{eqnarray}
\hat{g}_{h}\left(\theta\right)&=& \frac{1}{N-d-r+1}\sum^{N-d}_{n = r}\left[ \rho_{r}\left(h\right)-\frac{\left( \left(G^{\left(r\right)}_{n+h}\right)^2-\hat{\mu}_{r}\right)\left( \left(G^{\left(r\right)}_{n}\right)^2-\hat{\mu}_{r}\right)}{\hat{\gamma}_{r}\left(0\right)} \right]\nonumber\\
&=& \rho_{r}\left(h\right)- \hat{\rho}_{r}\left(h\right), \ \ h = 1, \ldots, d.\nonumber\\
\label{eq:empMoMconds}
\end{eqnarray}
The vector $\theta=\left(\mathbf{a},\mathbf{b}\right)$ containing the COGARCH(p,q) parameters are obtained by minimizing some distances between empirical and theoretical autocorrelations. The optimization problem is:
\begin{equation*}
\begin{array}{c}
\underset{
\theta \in \mathbb{R}^{q+p}
}{\min} d\left(\rho_{r},\hat{\rho}_{r}\right)
\end{array}
\end{equation*}
where the components of vectors $\rho_{r}$ and $\hat{\rho}_{r}$ are the theoretical and empirical autocorrelations defined in \eqref{COGARCH:autocorr} and \eqref{eq:empcorr} respectively. Function $d\left(x,y\right)$ measures the distance between vectors $x$ and $y$. In the \texttt{yuima} environment, three distances are available and listed  below:
\begin{enumerate}
\item the $L_1$ norm
\begin{equation}
\lVert \hat{g}\left(\theta\right)  \rVert_{1}=\sum_{h=1}^{d}\left| \hat{g}_{h}\left(\theta\right) \right|.
\label{eq:L1}
\end{equation}
\item the squared of $L_2$ norm 
\begin{equation}
\lVert \hat{g}\left(\theta\right) \rVert^{2}_{2}=\sum_{h=1}^{d}\left( \hat{g}_{h}\left(\theta\right) \right)^2.
\label{eq:L2}
\end{equation}
\item The quadratic form
\begin{equation}
\lVert \hat{g}\left(\theta\right) \rVert^{2}_{\mathbf{W}}= \hat{g}\left(\theta\right)^{\intercal}\mathbf{W}\hat{g}\left(\theta\right)
\label{eq:L2W}
\end{equation}
where the positive definite weighting matrix $\mathbf{W}$ is choosen to obtain efficient estimators between those that belong to the class of asymptotically normal estimators.
\end{enumerate}
It is worth noting that the objective function $\lVert \hat{g}\left(\theta\right) \rVert^{2}_{2}$ is a special case of the function $\lVert \hat{g}\left(\theta\right) \rVert^{2}_{\mathbf{W}}$ where the weighting matrix $\mathbf{W}$  coincides with the identity matrix. Both distances are related with the Generalized Method of Moments (GMM) introduced by \cite{Hansen1982}. Under some regularity conditions \cite{newey1994large}, the GMM estimators are consistent and, for    
 any general positive definite matrix $\mathbf{W}$, their asymptotic variance-covariance matrix $\mathbf{V}$ are:
\begin{equation*}
\mathbf{V} = \frac{1}{N-d-r+1}\left(\mathbf{D}^{\intercal}\mathbf{W}\mathbf{D}\right)^{-1} \mathbf{D}^{\intercal}\mathbf{W}\mathbf{S}\mathbf{W}\mathbf{D} \left(\mathbf{D}^{\intercal}\mathbf{W}\mathbf{D}\right)^{-1},
\end{equation*}
The matrix $\mathbf{D}$ is defined as:
\begin{equation}
\mathbf{D} = E\left[\frac{\partial f\left(G^{\left(r\right)}_{n},\mathbf{\theta}\right)  }{\partial \theta^{\intercal} }\right].
\end{equation}
While
\begin{equation}
\mathbf{S}=E\left[f\left(G^{\left(r\right)}_{n},\mathbf{\theta}\right) f\left( G^{\left(r\right)}_{n},\mathbf{\theta}\right)^{\intercal}\right].
\end{equation}
For the squared $L_2$ norm in \eqref{eq:L2} matrix $\mathbf{V}$ becomes:
\begin{equation}
\mathbf{V} = \frac{1}{N-d-r+1}\left(\mathbf{D}^{\intercal}\mathbf{D}\right)^{-1} \mathbf{D}^{\intercal}\mathbf{S}\mathbf{D} \left(\mathbf{D}^{\intercal}\mathbf{D}\right)^{-1}
\label{eqVarCov}
\end{equation}
while for \eqref{eq:L2W}, as observed above, the choice of the matrix $\mathbf{W}$ is done in order to obtain efficient estimators in the class of all asymptotically normal estimators. To obtain this result we prefer to use the Continuously Updated GMM estimator \cite{Hansen96}. In this case the matrix $\mathbf{W}$ is determinated simultaneusly with the estimation of the vector parameters $\theta$. Introducing the function $\lVert \hat{{g}}\left(\theta\right) \rVert^{2}_{\hat{\mathbf{W}}}$ as the sample counterpart of the quadratic form in \eqref{eq:L2W}, the minimization problem becomes:
\begin{equation*}
\begin{array}{c}
\underset{\theta \in \mathbb{R}^{q+p}}{\min}{\lVert {g}\left(\theta\right) \rVert^{2}_{\hat{\mathbf{W}}}= \hat{g}\left(\theta\right)^{\intercal}\hat{\mathbf{W}}\left(\theta\right)\hat{g}\left(\theta\right)}\\
\end{array}
\end{equation*}
where  function $\hat{\mathbf{W}}\left(\theta\right)$ maps from $\mathcal{R}^{p+q}$ to $\mathcal{R}^{d \times d}$ and is defined as:
\begin{equation}
\hat{\mathbf{W}}\left(\theta\right)=\left(\frac{1}{N-r-d+1}\sum_{n=r}^{N-r-d}f\left( G^{\left(r\right)}_{n},\theta\right)f\left( G^{\left(r\right)}_{n},\theta\right)^{\intercal}\right)^{-1}.
\end{equation}
Observe that $\hat{\mathbf{W}}\left(\theta\right)$ is a consistent estimator of matrix $\mathbf{S}^{-1}$ that means:
\begin{equation}
\hat{\mathbf{W}}\left(\theta\right) \underset{N \rightarrow +\infty}{\overset{P}{\rightarrow}} \mathbf{S}^{-1}
\end{equation}
consequently the asymptotic variance-covariance matrix $\mathbf{V}$ in $\eqref{eqVarCov}$ becomes:
\begin{equation}
\mathbf{V}=\frac{1}{N-r-d+1}\left(\mathbf{D}^{\intercal}\mathbf{S}^{-1}\mathbf{D}\right)^{-1}
\end{equation}

Once the estimates of vector $\theta$ are obtained, the user is allowed to retrieve the increments of the underlying L\'evy process according the following procedure. This stage is independent on the nature of the L\'evy measure but it is only based on the solution of the state process $Y_t$ and on the approximation of the quadratic variation with the squared increments of the L\'evy driven process. 

Starting from the discrete time equally spaced observations $G_{1},G_{2},\ldots,G_{N \Delta t}$, we remark that the increment $\Delta G_{t}:=G_{t}-G_{t-}$ can be approximated using the observations $\left\{G_{n \Delta t}\right\}_{n=0}^{N}$ as follows:
\begin{equation}
\Delta G_{t} \approx \Delta G_{n \Delta t} = G_{n \Delta t} -  G_{\left(n-1\right) \Delta t}
\label{incr:ApproxForRetriev}
\end{equation}
Recalling that $G_{t} = \sum_{0\leq s \leq t} \sqrt{V_{s}}\left(\Delta L_{s}\right)$, the approximation in \eqref{incr:ApproxForRetriev} becomes:
\begin{equation}
\Delta G_{t} \approx \sqrt{V_{n \Delta t}}\left(\Delta L_{n \Delta t}\right)
\label{approx:G_T}
\end{equation}
 where $V_{n \Delta t}$ is the value of the variance process at the observation time $t=n \frac{T}{N}$ and $\Delta L_{n \Delta t}= L_{n \Delta t}-L_{\left(n-1\right)\Delta t}$ is the discretized L\'evy increment from $t-\Delta t$ to $t$. \newline Using the discretization scheme introduced in  \eqref{apprMixed0} the process $Y_{n \delta t}$ is written as: 
\begin{equation}
Y_{n \delta t}\approx e^{A\Delta t}Y_{\left(n-1\right) \delta t}+e^{A\left(\Delta t\right)}e\left(a_{0}+a'Y_{t-\Delta t}\right)\left(\left[LL\right]_{n}^{d}-\left[LL\right]_{\left(n-1\right)}^{d}\right)
\label{scheme:approxYn}
\end{equation}
since the difference $\left[LL\right]_{n}^{d}-\left[LL\right]_{\left(n-1\right)}^{d}\approx \left(\Delta L_{n \Delta t}\right)^2$ and using the result in \eqref{approx:G_T}, the difference equation \eqref{scheme:approxYn} can be approximated in terms of the squared increments of COGARCH(p,q) process and we have:
\begin{equation}
\begin{array}{c}
Y_{t}\approx e^{A\Delta t}Y_{t-\Delta t}+e^{A\left(\Delta t\right)}e\left(a_{0}+a'Y_{t-\Delta t}\right)\left(\Delta L_{t}\right)^{2}\\
=e^{A\Delta t}Y_{t-\Delta t}+e^{A\left(\Delta t\right)}e\left(\Delta G_{t}\right)^{2}.
\end{array}
\label{mainResultForRecovNoise}
\end{equation}
Choosing $Y_0$ equal to the unconditional mean of the process $Y_{t}$, we are able to retrieve its sample path according to the recursive equation in \eqref{mainResultForRecovNoise}. The only quantities that we need to compute are the squared increments of the COGARCH(p,q) process on the grid $\left\{0, \Delta t,2\Delta t, \ldots, n\Delta t \ldots, N \Delta t\right\}$. The estimated state process in \eqref{mainResultForRecovNoise} is also useful for getting the estimated trajectory of the variance process. Finally note the L\'evy increment at  a general time $t=n  \Delta t$ is obtained as:   
\begin{equation}
\Delta L_{n\delta t}=\frac{\Delta G_{n \delta t}}{\sqrt{V_{n \delta t}}}.
\end{equation} 
Once the increments of the underlying L\'evy are obtained, it is possible to obtain the estimates for the L\'evy measure parameters through the Maximum Likelihood Estimation procedure [we refere to the yuima documentation \cite{Brousteetal2013, yuimaPack} for the available L\'evy processes].

\section{Package R}
\label{PackageRDescrip}
In this Section we illustrate the main classes and methods in \texttt{yuima} that allow the user to deal with a COGARCH(p,q) model. The  routines implemented are based on the results considered in Section \ref{SimulatCogarch} for simulation and Section \ref{TwoStepEstCOGARCH} for the estimation procedures.\newline
In particular we classify these classes and methods in three groups. The first group contains the classes and functions that allow the user to define a specific COGARCH(p,q) model in the \texttt{yuima} framework. The second group is used for the simulation of the sample paths for the COGARCH(p,q) model and the third is related to the possibility of estimation using simulated or real data.
\subsection{Classes and Methods for the definition of a COGARCH(P,Q) model}
The main object for a COGARCH(p,q) process in the \texttt{yuima} environment is an object of \texttt{yuima.cogarch-class} that contains all the information about a specific COGARCH(p,q) process. This class extends the \texttt{yuima.model-class} and it has only one additional slot, called $\texttt{@info}$, that contains an object of \texttt{cogarch.info-class}. We refer to the \texttt{yuima} documentation for a complete description of the slots that constitute the objects of class \texttt{yuima.model}. In this paper we focus only on the object of class \texttt{cogarch.info}.  In particular its structure is composed by the slots listed below:
\begin{itemize}
\item \texttt{@p} is an integer that is the number of moving average coefficients in the Variance process $V_{t}$.
\item \texttt{@q} is an  integer number that corresponds to the dimension of autoregressive coefficients in the variance process $V_{t}$. 
\item \texttt{@ar.par} contains a string that is the label for the autoregressive coefficients.
\item \texttt{@ma.par} is the  Label for the moving average coefficients.
\item \texttt{@loc.par} indicates the name of the location coefficient in the process $V_t$.
\item \texttt{@Cogarch.var} string that contains the name of the observed process $G_{t}$.
\item \texttt{@V.var} is the Label of the $V_{t}$ process.
\item \texttt{@Latent.var} indicates the label of the state process $Y_t$.
\item \texttt{@XinExpr} is a logical variable. If the value is \texttt{FALSE}, default value, the starting condition for the state process $Y_t$  is a zero vector. Otherwise the user has  to fix the starting condition as argument \texttt{true.parameter} in the method \texttt{simulate}. 
\item \texttt{@measure} identifies the L\'evy measure of the underlying noise and consequently the discrete part of the quadratic variation that drives the state process.
\item \texttt{@measure.type} says if the L\'evy measure belongs to the family of Compound Poisson or is another type of L\'evy
\end{itemize}
The user builds an object of class \texttt{yuima.cogarch} through  to the constructor \texttt{setCogarch}:

\texttt{
setCogarch(p, q, ar.par = "b", ma.par = "a", loc.par = "a0", Cogarch.var = "g", V.var = "v", 
           Latent.var = "y", jump.variable = "z",  time.variable = "t", measure = NULL, 
           measure.type = NULL, XinExpr = FALSE, startCogarch = 0, work = FALSE, ...)
}

The arguments used in a call of the function \texttt{setCogarch} are illustrated in the following list:

\begin{itemize}
  \item \texttt{p}: A non negative integer that is the number of the moving average coefficients in the variance process.
  \item  \texttt{q}:  A non negative integer that indicates the number of the autoregressive  coefficients in the variance process. 
  \item  \texttt{ar.par}: A character-string that is the label of the autoregressive coefficients.
  \item  \texttt{ma.par}: A character-string that is the label of the autoregressive coefficients.
  \item  \texttt{loc.par}: A string that indicates the label of the location coefficient in the variance process.
  \item  \texttt{Cogarch.var}: A character-string that is the label of the observed COGARCH process.
  \item \texttt{V.var}: A character-string that is the label of the latent variance process.
  \item \texttt{Latent.var}:  A character-string that is the label of the latent process in the state space representation for the variance process.
  \item \texttt{jump.variable}: Label of the underlying L\'evy process .
  \item \texttt{time.variable}: Label of the time variable.
  \item  \texttt{measure}: L\'evy measure of the underlying L\'evy process.
  \item \texttt{measure.type}: Label that indicates whether the underlying noise is a Compound Poisson process or another L\'evy without the diffusion component.
  \item \texttt{XinExpr}: A logical variable that identifies the starting condition. In particular, the default value \texttt{XinExpr = FALSE} implies that the starting
condition for the state process is zero. Alternatively \texttt{XinExpr = TRUE} means that the user is allowed to specify as parameters the starting values for each component of the state variable. 
  \item \texttt{startCogarch}: Initial value for the COGARCH process.
  \item \texttt{$\dots$}: Arguments to be passed to \texttt{setCogarch} such as the slots of the \texttt{yuima.model-class}.
\end{itemize}

\subsection{Classes and Methods for the simulation of a COGARCH(P,Q) model}
The \texttt{simulate} is a method for an object of class \texttt{yuima.model}. It is also available for an object of class \texttt{yuima.cogarch}. The function requires the following inputs:

\texttt{
simulate(object, nsim=1, seed=NULL, xinit, true.parameter, space.discretized = FALSE, increment.W = NULL, 
         increment.L = NULL, method = "euler", hurst, methodfGn = "WoodChan", sampling=sampling, 
         subsampling=subsampling, ...)
}

In this work we focus on the argument \texttt{method} that identifies the type of discretization scheme for the time when the \texttt{object} belongs to the class of \texttt{yuima.cogarch}. The default value \texttt{euler} means that the simulation of a sample path is based on the Euler-Maruyama discretization of the stochastic differential equations. This approach is available for all objects of class \texttt{yuima.model}. For the COGARCH(p,q) an alternative simulation scheme is available choosing \texttt{method = mixed}. In this case the generation of trajectories is based on the solution \eqref{explicitSolu} for the state process. In particular if the underlying noise is a Compound Poisson L\'evy process, the trajectory is built using a two step algorithm. First the jump time is simulated internally using the Exponential distribution with parameter $\lambda$ and then the size of jump is simulated using the random variable specified in the slot \texttt{yuima.cogarch@model@measure}. For the other L\'evy processes, the simulation scheme is based on the discretization of the state process solution \eqref{apprMixed0} in Section \ref{PackageRDescrip}.       

\subsection{Classes and Methods for the estimation of a COGARCH(P,Q) model}

The \texttt{cogarch.gmm} class is a class of the  \texttt{yuima} package that contains estimated parameters obtained by the \texttt{gmm} function. 

\begin{itemize}
\item \texttt{@model} is an object of of  \texttt{yuima.cogarch-class}.
\item \texttt{@objFun} is an object of class \texttt{character} that indicates the objective function used in the minimization problem. \texttt{L2} refers to  the squared of $L_2$ norm in \eqref{eq:L2}, \texttt{L2CUE} for the quadratic form \eqref{eq:L2W} and the \texttt{L1} for the $L_1$ norm in \eqref{eq:L1}    
\item \texttt{@call} is an object of class \texttt{language}.
\item \texttt{@coef} is an object of class \texttt{numeric} that contains the estimated parameters.
\item \texttt{@fullcoef} is an object of class \texttt{numeric} that contains the estimated and fixed parameters from the user.
\item \texttt{@vcov} is an object of class \texttt{matrix}.
\item \texttt{@min} is an object of class \texttt{numeric}.
\item \texttt{@minuslogl} is an object of class \texttt{function}.
\item \texttt{@method} is an object of class \texttt{character}.
\end{itemize}

 The \texttt{cogarch.gmm.incr} is a class of the  \texttt{yuima} package that extends the \texttt{cogarch.gmm-class} and is filled from the function \texttt{gmm}.

\begin{itemize}
\item \texttt{Incr.Lev} is an object of class \texttt{zoo} that contains the estimated increments of the noise obtained using \texttt{cogarchNoise}.
\item \texttt{model}is an object of \texttt{yuima.cogarch-class}.
    \item \texttt{logL.Incr} is an object of class \texttt{numeric} that contains the value of the log-likelihood for the estimated Levy increments.
    \item \texttt{objFun} is an object of class \texttt{character} that indicates the objective function used in the minimization problem. The values are the same for the slot \texttt{@objFun} in an object of class \texttt{cogarch.gmm}.
    \item \texttt{call} is an object of class \texttt{language}. 
    \item \texttt{coef} is an object of class \texttt{numeric} that contains the estimated parameters.
    \item \texttt{fullcoef} is an object of class \texttt{numeric} that contains estimated and fixed parameters.
    \item \texttt{vcov} is an object of class \texttt{matrix}.
    \item \texttt{min} is an object of class \texttt{numeric}.
    \item \texttt{minuslogl} is an object of class \texttt{function}.
    \item \texttt{method} is an object of class \texttt{character}.
\end{itemize}

The function \texttt{gmm} returns the estimated parameters of a COGARCH(p,q) model. The parameters are obtained by matching the theoretical with the empirical autocorrelation function. 

\texttt{
gmm(yuima, data = NULL, start, method="BFGS", fixed = list(), lower, upper, 
    lag.max = NULL, equally.spaced = TRUE, Est.Incr = "NoIncr", objFun = "L2")
}

\begin{itemize}
\item \texttt{yuima} is a \texttt{yuima} object or an object of \texttt{yuima.cogarch-class}
  \item \texttt{data} is an object of class \texttt{yuima.data-class} contains the observations available at uniformly spaced instants of time. If \texttt{data=NULL}, the default, the function uses the data in an object of \texttt{yuima-class}.
  \item \texttt{start} is a \texttt{list} that contains the starting values for the optimization routine.
  \item \texttt{method} is a string that indicates one of the methods available in the function \texttt{optim}.
  \item \texttt{fixed} a list of fixed parameters in the optimization routine.
  \item \texttt{lower} a named list for specifying lower bounds for parameters.
  \item \texttt{upper} a named list for specifying upper bounds for parameters.
  \item \texttt{lag.max} maximum lag for which we calculate the theoretical and empirical acf. Default is $\sqrt{N}$ where \texttt{N} is the number of observations.
  \item \texttt{equally.spaced} Logical variable. If \texttt{equally.spaced = TRUE}, in each unitary interval we have the some number of observations. If \texttt{equally.spaced = FALSE}, each unitary interval is composed by different number of observations.
  \item \texttt{Est.Incr}  a string variable. If \texttt{Est.Incr = "NoIncr"}, default value, \texttt{gmm} returns an object of class  \texttt{cogarch.gmm-class} that contains the COGARCH parameters. 
  If \texttt{Est.Incr = "Incr"} or \texttt{Est.Incr = "IncrPar"} the output is an object of class \texttt{cogarch.gmm.incr-class}. In the first case the object contains the increments of the underlying noise while in the second case it contains also the estimated parameters of the L\'evy measure.
  \item \texttt{objFun} a string variable that indentifies the objective function in the optimization step. \texttt{objFun = "L2"}, default value, the objective function is  a quadratic form where the weighting matrix is the identity one. \texttt{objFun = "L2CUE"} the weighting matrix is estimated using Continuously Updating GMM (L2CUE). 
  \texttt{objFun = "L1"}, the objective function is the mean absolute error. In the last case standard errors for estimators are not available.
  \end{itemize}
  
  Function \texttt{gmm} uses function \texttt{cogarchNoise} for the estimation of the underlying L\'evy in a COGARCH(p,q) model. This function assumes that the underlying L\'evy process is symmetric and centered in zero.

\texttt{
cogarchNoise(yuima.cogarch, data=NULL, param, mu=1)
}

The arguments of the \texttt{cogarchNoise} are listed below

\begin{itemize}
  \item \texttt{yuima.cogarch}  is an object of \texttt{yuima-class} or an object of \texttt{yuima.cogarch-class} that contains all the information about the COGARCH(p,q) model. 
  \item \texttt{data} is an object of class \texttt{yuima.data-class} that contains the observations available at uniformly spaced instants of time. If \texttt{data=NULL}, the default, the \texttt{cogarchNoise} uses the data in an object of \texttt{yuima.data-class}.
  \item \texttt{param} is a \texttt{list} of parameters for the COGARCH(p,q) model.
  \item \texttt{mu} is a \texttt{numeric} object that contains the value of the second moments of the L\'evy measure.
\end{itemize}


\section{Numerical results}
\label{NumExamp}
In this section we show how to use the \texttt{yuima} package for the simulation and the estimation of a COGARCH(p,q) model driven by  different symmetric L\'evy processes. As a first step we focus on a COGARCH(1,1) model driven by different L\'evy processes available on the package. In particular we consider the cases in which the driven noise are a Compound Poisson with jump size normally distributed and a Variance Gamma process. In the last part of this section, we show also that the estimation procedure implemented seems to be adequate even for higher order COGARCH models. In particular we simulate and then estimate different kind of COGARCH(p,q) models driven by a Compound Poisson process where the distribution of the jump size is a normal.

\subsection{Simulation and Estimation of a COGARCH(1,1)}

The first example is a COGARCH(1,1) model driven by a Compound Poisson process. As a first step, we choose the set of the model parameters:
\begin{verbatim}
> numSeed <- 200
> param.cp <- list(a1 = 0.038, b1 =  0.053, a0 = 0.04/0.053, 
+             lambda = 1, eta=0, sig2=1, x01 = 50.33)
\end{verbatim}
$a_{1}, b_1$ and $a_0$ are the parameters of the state process $Y_{t}$. $\lambda$ is the intensity of the Compound Poisson process while $\eta$ and $\sigma^2$ are the mean and the variance  of the jump size. $x_{0,1}$ is the starting point of the process $X_{t}$, the choosen value is the stationary mean of the state process and it is used in the simulation algorithm.\newline In the following command line we define the model using the \texttt{setCogarch} function.
\begin{verbatim}
> mod.cp <- setCogarch(p = 1, q = 1, work = FALSE,
+           measure=list(intensity="lambda",df=list("dnorm(z,eta,sig2)")),measure.type = "CP", 
+           Cogarch.var = "g", V.var = "v", Latent.var="x",
+           XinExpr=TRUE)
\end{verbatim}
We simulate a sample path of the model using the Euler discretization. We fix $\Delta t =\frac{1}{15}$ and the command lines below are used to instruct \texttt{yuima} for the choice of the simulation scheme:
\begin{verbatim}
> Term <- 1600
> num <- 24000
> set.seed(numSeed)
> samp.cp <- setSampling(Terminal=Term, n=num)
> sim.cp <- simulate(mod.cp, true.parameter=param.cp,
+           sampling=samp.cp, method="euler")
\end{verbatim}
In the following figure we show the behaviour of the simulated trajectories for the COGARCH(1,1)  model $G_t$, the variance $V_t$ and the state space $X_t$:
\begin{verbatim}
> plot(sim.cp, main = "simulated COGARCH(1,1) model driven by a Compound Poisson process")
\end{verbatim}

\begin{figure}[h]
	\centering
		\includegraphics[width=0.50\textwidth]{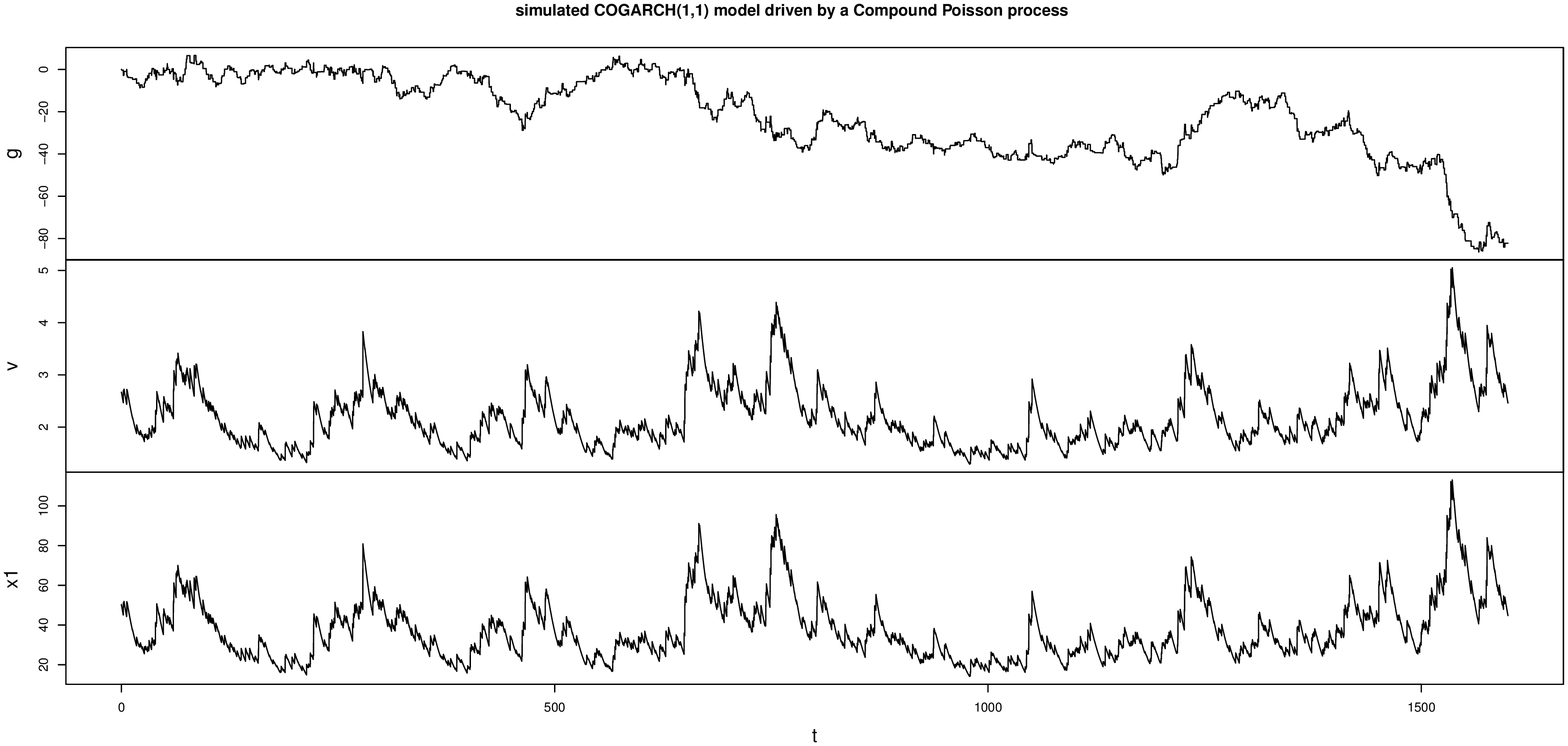}
\end{figure}

We use the two step algorithm developed in Section \ref{TwoStepEstCOGARCH} for the estimation of the COGARCH(p,q) and the L\'evy measure parameters. In the \texttt{yuima} function \texttt{gmm}, we fix \texttt{objFun = L2} meaning that the objective function used in the minimization is the mean squared error. Setting also \texttt{Est.Incr=IncrPar}, the function \texttt{gmm} returns the estimated increments of the underlying noise. 
\begin{verbatim}
> res.cp <- gmm(sim.cp, start = param.cp, objFun = "L2", Est.Incr = "IncrPar")
\end{verbatim}
The results can be displayed using the method \texttt{summary} and in the following figure we report the recovered increments of the underlying noice process.
\begin{verbatim}
> summary(res.cp)

Two Stages GMM estimation 

Call:
gmm(yuima = sim.cp, start = param.cp, Est.Incr = "IncrPar", objFun = "L2")

Coefficients:
           Estimate Std. Error
b1     6.783324e-02 0.06862392
a1     3.403071e-02 0.02897625
a0     1.032014e+00         NA
lambda 1.073912e+00         NA
eta    6.818470e-09         NA
sig2   7.837838e-01         NA

 Log.objFun L2: -3.491179 


Number of increments: 24000

Average of increments: -0.002114

Standard Dev. of increments: 0.256610


-2 log L of increments: 2851.529874

Summary statistics for increments:
     Min.   1st Qu.    Median      Mean   3rd Qu.      Max. 
-2.840000  0.000000  0.000000 -0.002114  0.000000  3.686000 
\end{verbatim}
\begin{verbatim}
> plot(res.cp, main = "Compound Poisson Increment of a COGARCH(1,1) model")
\end{verbatim}

\begin{figure}[h]
	\centering
		\includegraphics[width=0.50\textwidth]{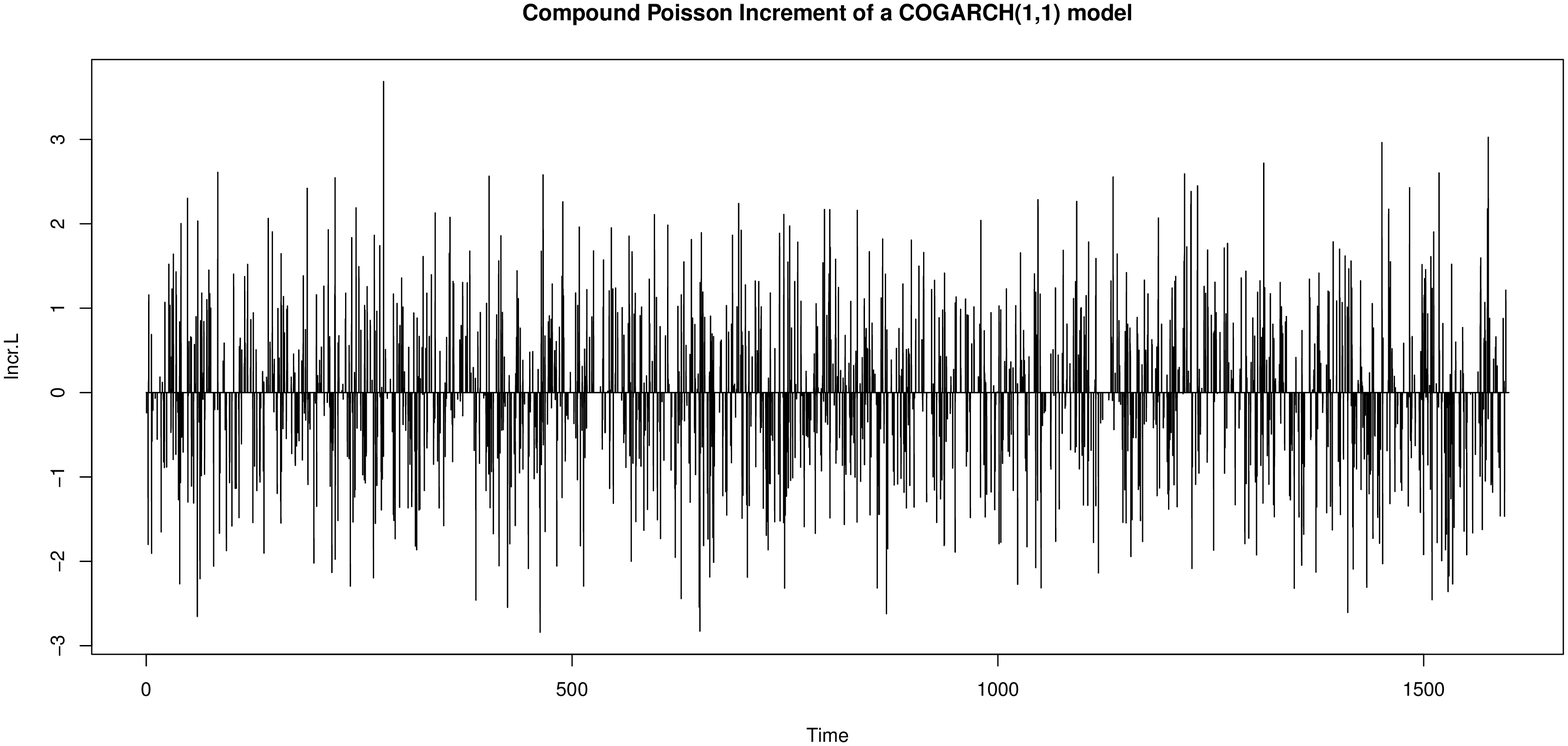}
\end{figure}

We are able also to generate the original process using the increments stored into the object \texttt{res.cp} using the \texttt{simulate} function.
\begin{verbatim}
> traj.cp<- simulate(res.cp)
> plot(traj.cp, main = "estimated COGARCH(1,1) driven by compound poisson process")
\end{verbatim}

\begin{figure}[h]
	\centering
		\includegraphics[width=0.50\textwidth]{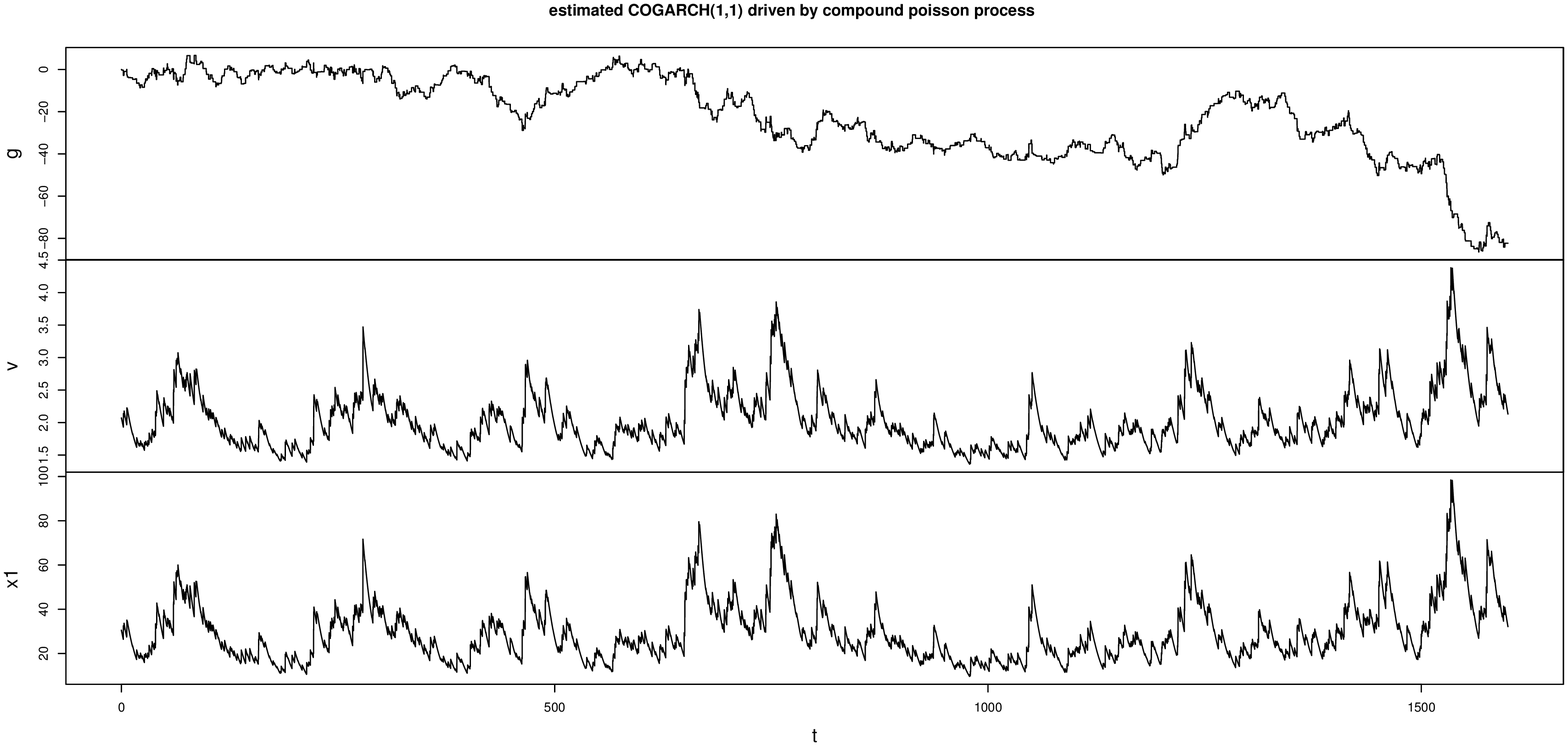}
\end{figure}

In the next example, we simulate and estimate a COGARCH(1,1) model driven by a Variance Gamma process. We set the values for the parameters and define the model using the following command lines:
\begin{verbatim}
> param.VG <- list(a1 = 0.038,  b1 =  0.053, a0 = 0.04/0.053,
+             lambda = 1, alpha = sqrt(2), beta = 0, mu = 0, x01 = 50.33)
> cog.VG <- setCogarch(p = 1, q = 1, work = FALSE,
+           measure=list("rngamma(z, lambda, alpha, beta, mu)"),
+           measure.type = "code", Cogarch.var = "y", V.var = "v",
+           Latent.var = "x", XinExpr = TRUE)
\end{verbatim}
We obtain a trajectory for COGARCH(1,1) with Variance Gamma noise.
\begin{verbatim}
> set.seed(numSeed)
> samp.VG <- setSampling(Terminal = Term, n = num)
> sim.VG <- simulate(cog.VG, true.parameter = param.VG,
+           sampling = samp.VG, method = "euler")
> plot(sim.VG, main = "simulated COGARCH(1,1) model driven by a Variance Gamma process")
\end{verbatim}

\begin{figure}[h]
	\centering
		\includegraphics[width=0.50\textwidth]{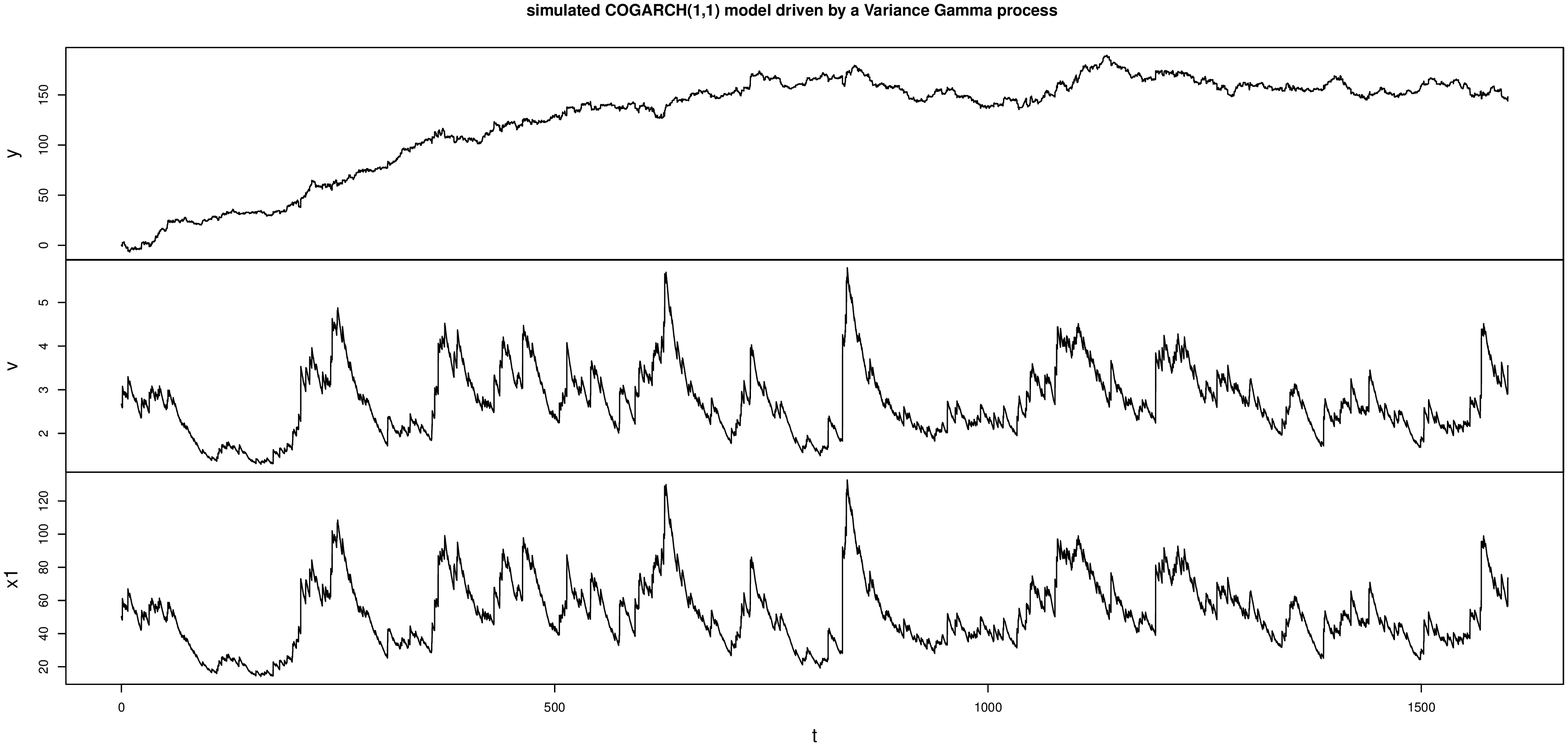}
\end{figure}

and then we estimate the model parameters:

\begin{verbatim}
> res.VG <- gmm(sim.VG, start = param.VG, Est.Incr = "IncrPar")
> summary(res.VG)

Two Stages GMM estimation 

Call:
gmm(yuima = sim.VG, start = param.VG, Est.Incr = "IncrPar")

Coefficients:
          Estimate Std. Error
b1     0.051449188 0.04168365
a1     0.028791052 0.01810412
a0     1.248576654         NA
lambda 1.049274382 0.09432438
alpha  1.466220182 0.08769087
beta   0.051526860 0.03929050
mu     0.003357025 0.02935151

 Log.objFun L2: -3.755496 


Number of increments: 24000

Average of increments: 0.003635

Standard Dev. of increments: 0.258127


-2 log L of increments: 4291.499302

Summary statistics for increments:
     Min.   1st Qu.    Median      Mean   3rd Qu.      Max. 
-5.548000 -0.001729  0.000000  0.003635  0.002092  4.005000 
\end{verbatim}
\begin{verbatim}
> plot(res.VG, main = "Variance Gamma Increment of a COGARCH(2,1) model")
\end{verbatim}

\begin{figure}[h]
	\centering
		\includegraphics[width=0.50\textwidth]{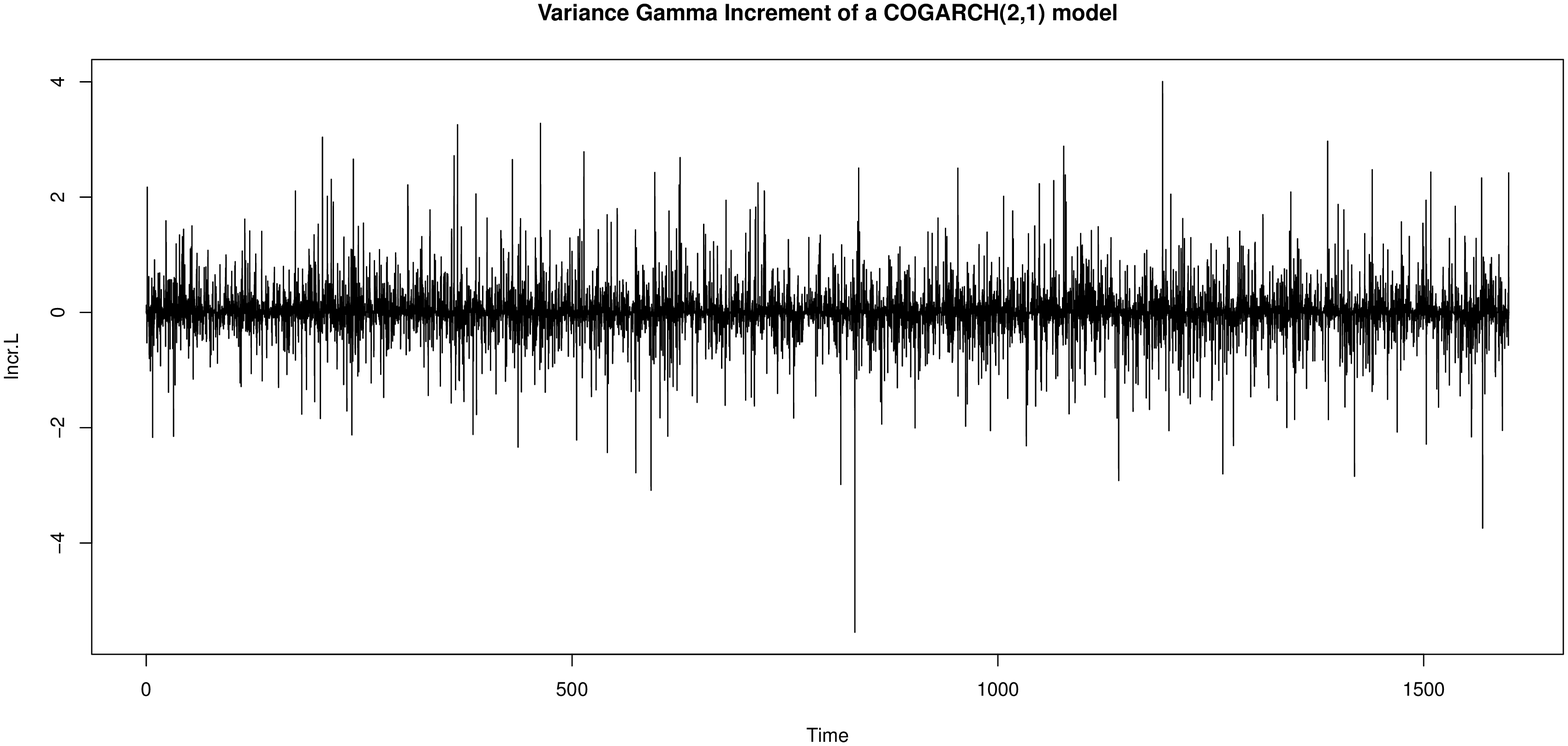}
\end{figure}

Even in this case we can obtain the COGARCH(1,1) trajectory using the estimated increments as follows:

\begin{verbatim}

> traj.VG <- simulate(res.VG)
> plot(traj.VG, main="estimated COGARCH(1,1) model driven by Variance Gamma process")
\end{verbatim}

\begin{figure}[h]
	\centering
		\includegraphics[width=0.50\textwidth]{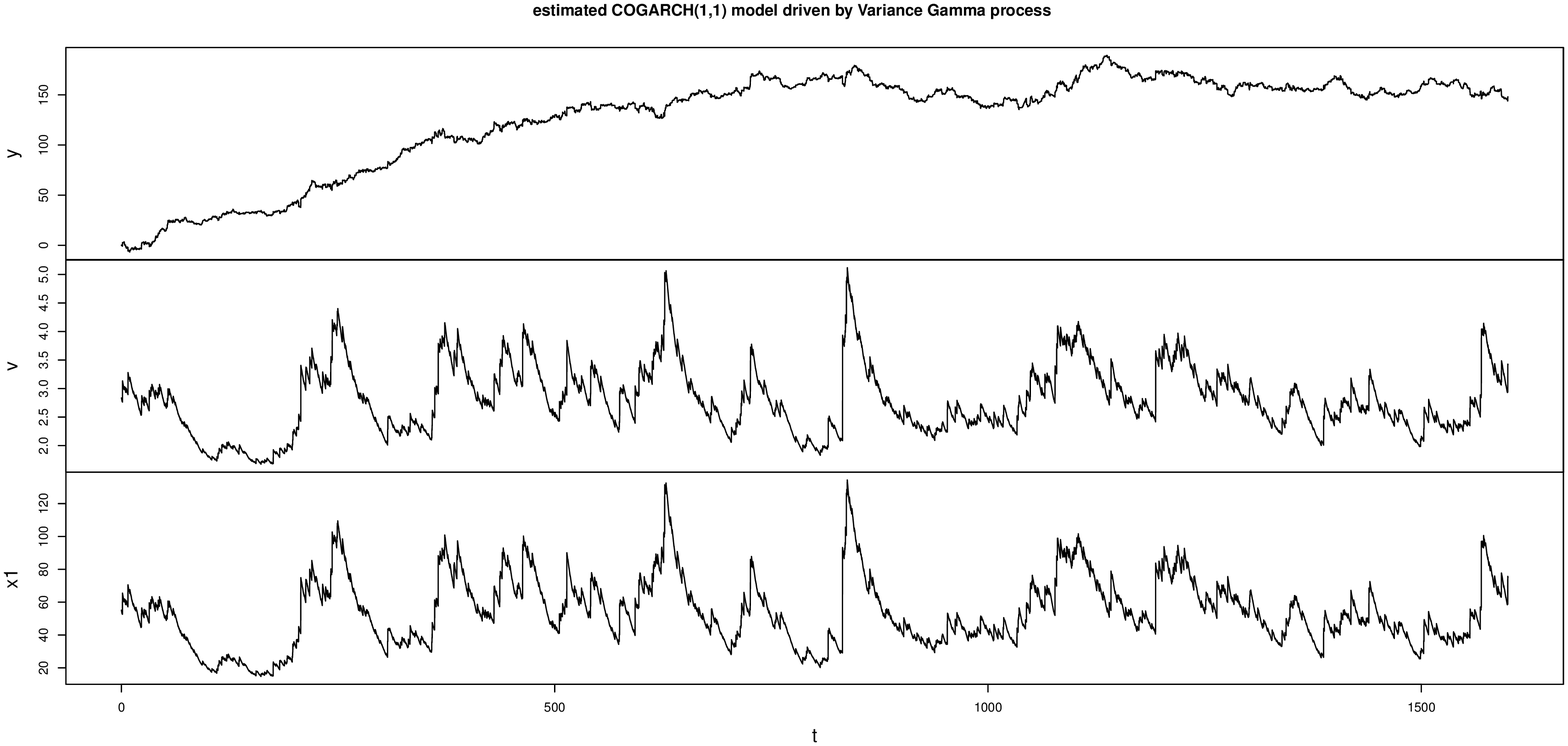}
\end{figure}

%
%
\subsection{COGARCH(p,q) model driven by a Compound Poisson process}
We conclude this Section by illustrating an example using COGARCH(p,q) process. In this way, we show the  ability of the \texttt{yuima} package in managing in a complete way these models. For this reason in the following we consider a COGARCH(2,1) 
 driven by Compound Poisson Processes where the jump size is normally distributed.\newline   
We define the COGARCH(2,1) model in the \texttt{yuima} using the command lines:
\begin{verbatim}
> param.cp2 <- list(a0 = 0.5, a1 = 0.1, b1 =1.5, b2 = 0.5, 
+              lambda = 1, eta = 0, sig2 = 1, x01 = 2.5, x02 = 0)
> mod.cp2 <- setCogarch(p = 1, q = 2, work = FALSE,
+            measure = list(intensity = "lambda",df = list("dnorm(z,eta,sig2)")),
+            measure.type = "CP", Cogarch.var = "y", V.var = "v",
+            Latent.var = "x", XinExpr = TRUE)
\end{verbatim}

We simulate a trajectory.

\begin{verbatim}
> samp.cp2 <- setSampling(Terminal = Term, n = num)
> set.seed(numSeed)
> sim.cp2 <- simulate(mod.cp2, sampling = samp.cp2, 
+            true.parameter = param.cp2, method="euler")
> plot(sim.cp2, main = "simulated COGARCH(2,1) model driven by a Compound Poisson process")
\end{verbatim}

\begin{figure}[h]
	\centering
		\includegraphics[width=0.50\textwidth]{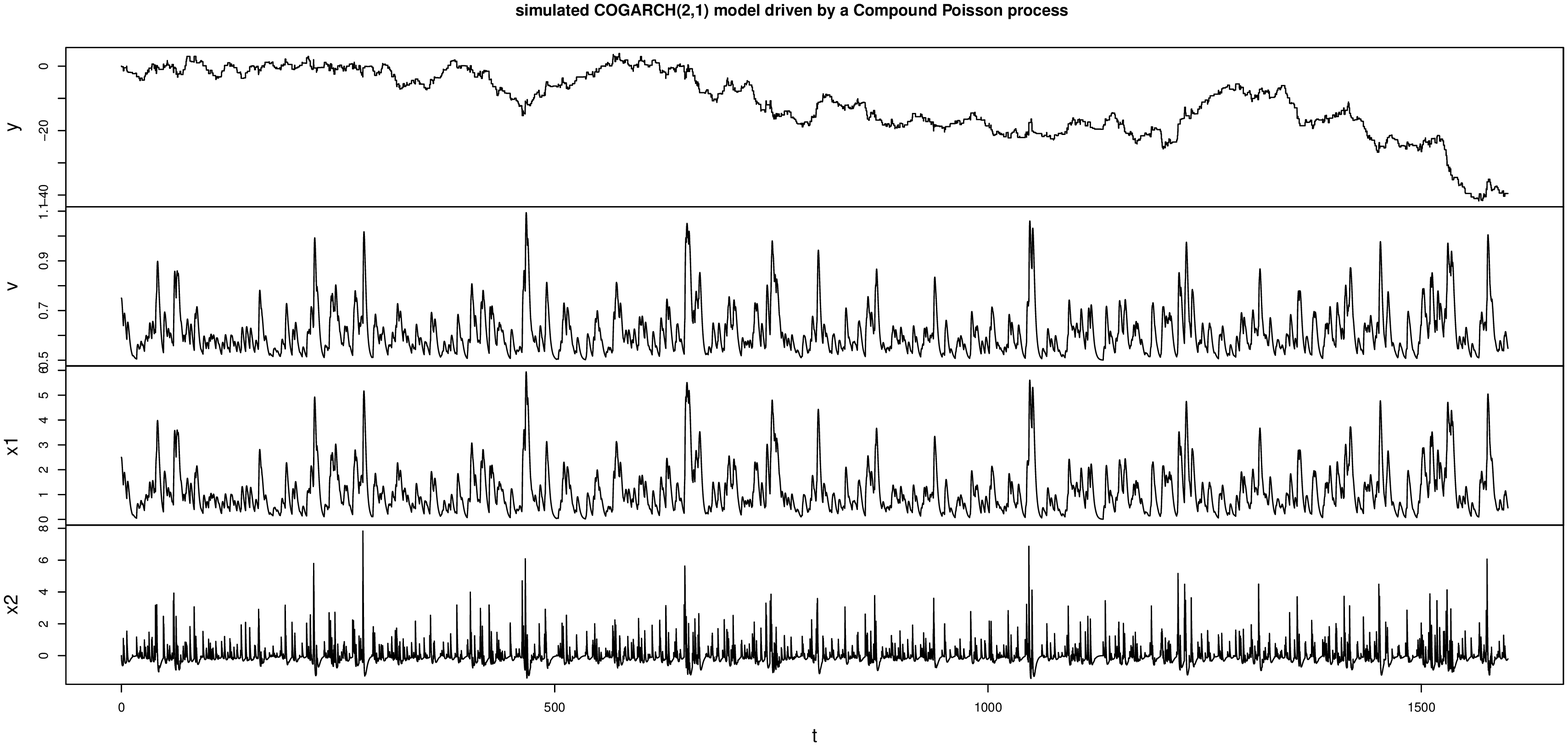}
\end{figure}

We estimate the model parameters and recover the underlying L\'evy noise increments:

\begin{verbatim}
> res.cp2 <- gmm(yuima = sim.cp2, start = param.cp2, Est.Incr = "IncrPar")
> summary(res.cp2)

Two Stages GMM estimation 

Call:
gmm(yuima = sim.cp2, start = param.cp2, Est.Incr = "IncrPar")

Coefficients:
           Estimate Std. Error
b2     0.0569630413 0.20054247
b1     0.9520642366 3.54500502
a1     0.0281299955 0.09775311
a0     0.2956658497         NA
lambda 1.0423762156         NA
eta    0.0002425553         NA
sig2   0.8154399532         NA

 Log.objFun L2: -3.323979 


Number of increments: 24000

Average of increments: -0.001929

Standard Dev. of increments: 0.258830


-2 log L of increments: 2861.417140

Summary statistics for increments:
     Min.   1st Qu.    Median      Mean   3rd Qu.      Max. 
-3.054000  0.000000  0.000000 -0.001929  0.000000  3.483000 
\end{verbatim}

\begin{verbatim}
> plot(res.cp2, main = "Compound Poisson Increment of a COGARCH(2,1) model")
\end{verbatim}

\begin{figure}[h]
	\centering
		\includegraphics[width=0.50\textwidth]{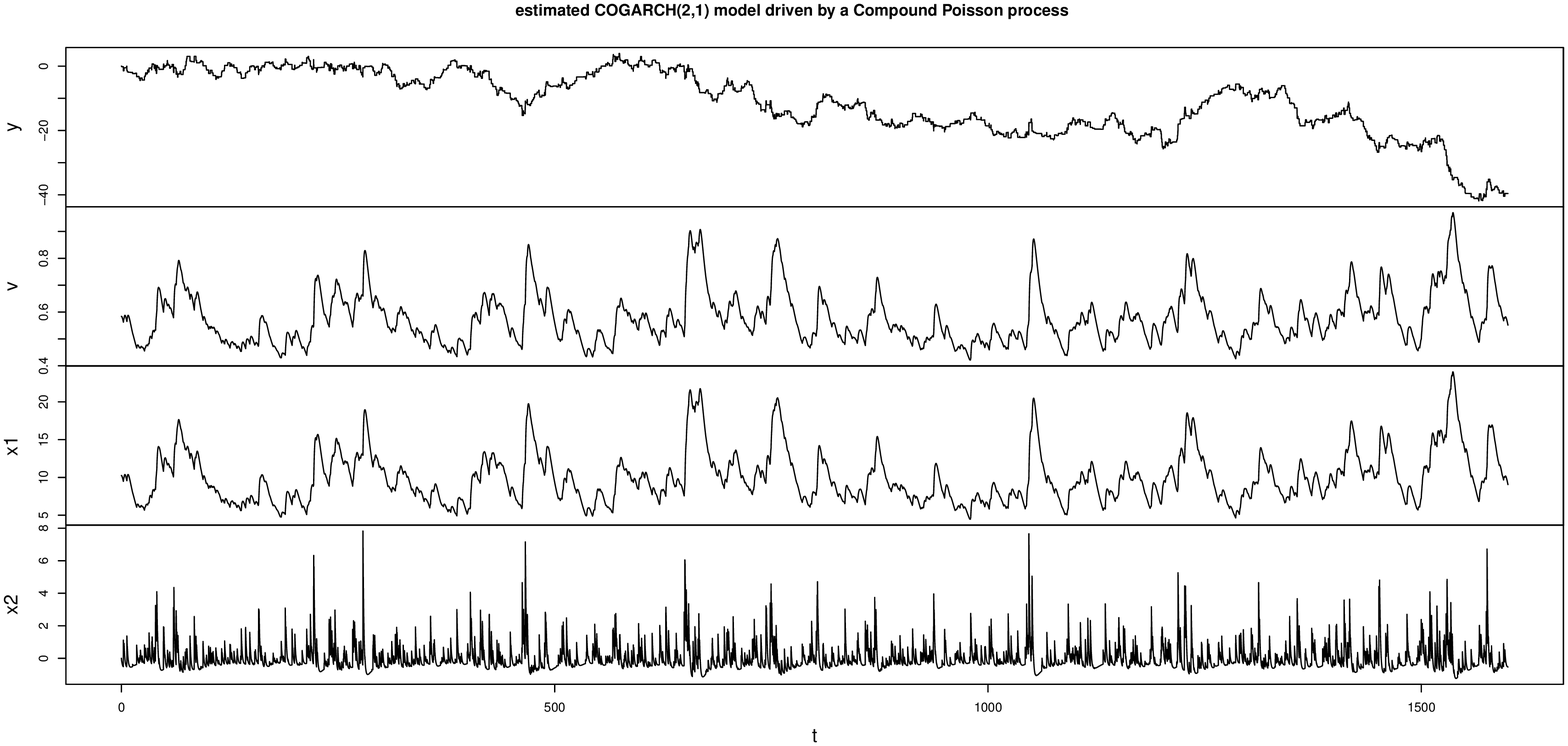}
		\end{figure}

The path of the COGARCH(2,1) driven by the estimated increments are reported below:

\begin{verbatim}
> traj.cp2 <- simulate(res.cp2)
> plot(traj.cp2, main = "estimated COGARCH(2,1) model driven by a Compound Poisson process")
\end{verbatim}

\begin{figure}[h]
	\centering
		\includegraphics[width=0.50\textwidth]{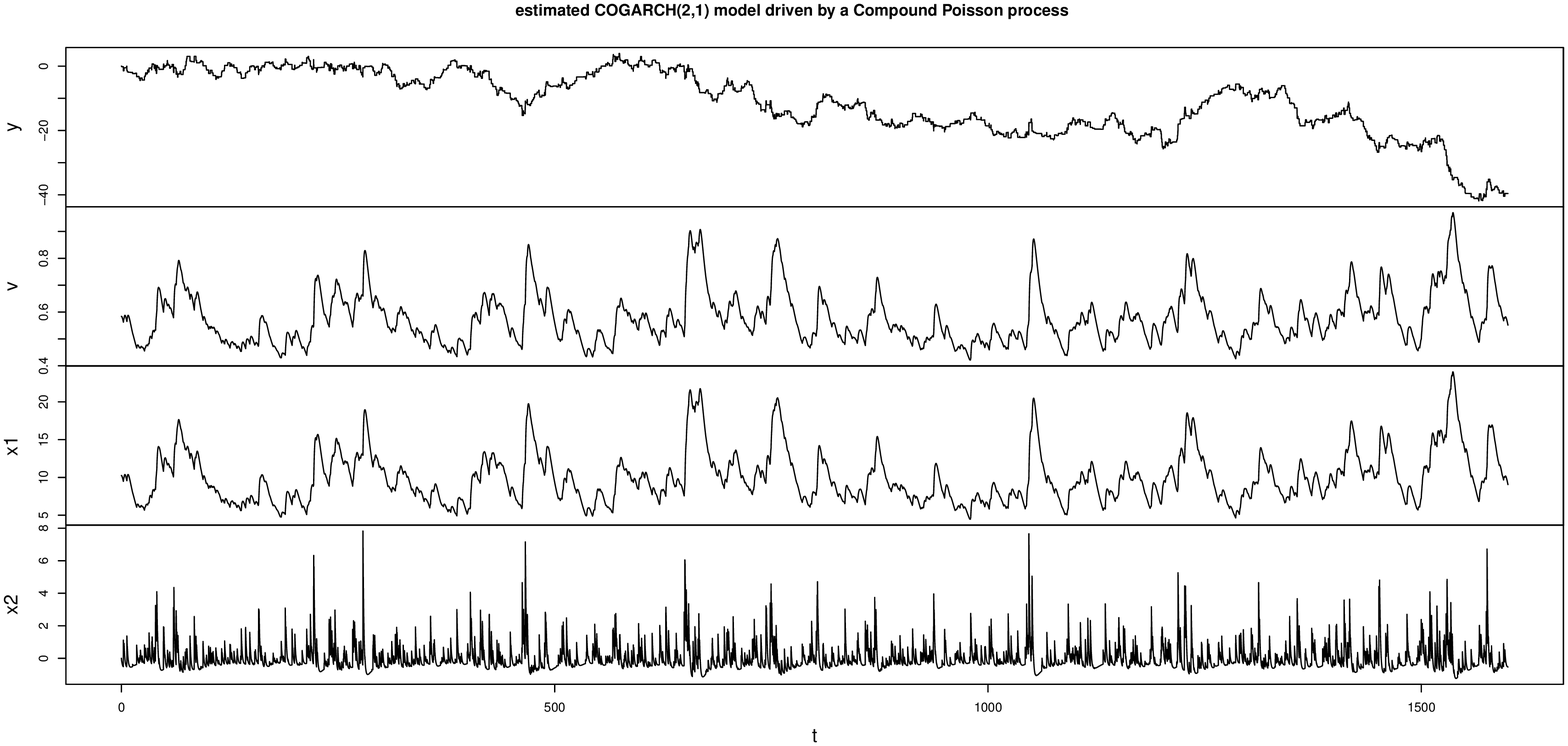}
		\end{figure}

\section*{Acknowledgements}
The authors would like to thank the CREST Japan Science and Technology Agency.

%
%

\bibliographystyle{plain}

\begin{thebibliography}{10}

\bibitem{Basrak2002}
B.~Basrak, R.~A. Davis, and T.~Mikosch.
\newblock A characterization of multivariate regular variation.
\newblock {\em The Annals of Applied Probability}, 12(3):908--920, 2002.

\bibitem{Bibbona2015}
E.~Bibbona and I.~Negri.
\newblock Higher moments and prediction-based estimation for the cogarch(1,1)
  model.
\newblock {\em Scandinavian Journal of Statistics}, pages n/a--n/a, 2015.

\bibitem{Bollerslev86}
T.~Bollerslev.
\newblock {Generalized autoregressive conditional heteroskedasticity}.
\newblock {\em Journal of Econometrics}, 31(3):307--327, April 1986.

\bibitem{Brandt1986}
A.~Brandt.
\newblock The stochastic equation yn+1=anyn+bn with stationary coefficients.
\newblock {\em Advances in Applied Probability}, 18(1):211--220, 1986.

\bibitem{Brockwell2006}
P.~Brockwell, E.~Chadraa, and A.~Lindner.
\newblock {Continuous-time GARCH processes}.
\newblock {\em Annals of Applied Probability}, 16(2):790--826, 2006.

\bibitem{BrockwellDavisYang2007}
P.~J. Brockwell, R.~A. Davis, and Y.~Yang.
\newblock Estimation for non-negative l\'{e}vy-driven ornstein-uhlenbeck
  processes.
\newblock {\em Journal of Applied Probability}, 44:987--989, 2007.

\bibitem{Brockwell2001}
P.J. Brockwell.
\newblock {L\'evy-driven carma processes}.
\newblock {\em Annals of the Institute of Statistical Mathematics},
  53(1):113--124, 2001.

\bibitem{Brousteetal2013}
A.~Brouste, M.~Fukasawa, H.~Hino, S.~M. Iacus, K.~Kamatani, Y.~Koike,
  H.~Masuda, R.~Nomura, T.~Ogihara, Shimuzu Y., M.~Uchida, and Yoshida N.
\newblock The yuima project: A computational framework for simulation and
  inference of stochastic differential equations.
\newblock {\em Journal of Statistical Software}, 57(4):1--51, 2014.

\bibitem{Chadraa2010Thesis}
E.~Chadraa.
\newblock {\em Statistical Modelling with COGARCH(P,Q) Processes.}, 2009.
\newblock PhD Thesis.

\bibitem{Cont01empiricalproperties}
R.~Cont.
\newblock Empirical properties of asset returns: stylized facts and statistical
  issues.
\newblock {\em Quantitative Finance}, 1:223--236, 2001.

\bibitem{Hansen1982}
L.~P. Hansen.
\newblock Large sample properties of generalized method of moments estimators.
\newblock {\em Econometrica}, 50(4):1029--1054, 1982.

\bibitem{Hansen96}
L.~P. Hansen, J.~Heaton, and A.~Yaron.
\newblock Finite-sample properties of some alternative gmm estimators.
\newblock {\em Journal of Business \& Economic Statistics}, 14(3):262--280,
  1996.

\bibitem{Haug2007}
S.~Haug, C.~Kl\"{u}ppelberg, A.~Lindner, and M.~Zapp.
\newblock Method of moment estimation in the cogarch(1,1) model.
\newblock {\em Econometrics Journal}, 10(2):320--341, 2007.

\bibitem{R:Iacus:2007}
S.~M. Iacus.
\newblock {\em Simulation and Inference for Stochastic Differential Equations:
  With R Examples}.
\newblock Springer, 2008.

\bibitem{IacusMercur2015}
S.~M. Iacus and L.~Mercuri.
\newblock Implementation of lévy carma model in yuima package.
\newblock {\em Computational Statistics}, pages 1--31, 2015.

\bibitem{Kallsen200974}
J.~Kallsen and B.~Vesenmayer.
\newblock \{COGARCH\} as a continuous-time limit of garch(1,1).
\newblock {\em Stochastic Processes and their Applications}, 119(1):74 -- 98,
  2009.

\bibitem{Kesten1973}
H.~Kesten.
\newblock Random difference equations and renewal theory for products of random
  matrices.
\newblock {\em Acta Mathematica}, 131(1):207--248, 1973.

\bibitem{Cogarch2004}
C.~Kl\"{u}ppelberg, A.~Lindner, and R.~Maller.
\newblock A continuous-time garch process driven by a l\'{e}vy process:
  Stationarity and second-order behaviour.
\newblock {\em Journal of Applied Probability}, 41(3):601--622, 2004.

\bibitem{Loregian2012}
A.~Loregian, L.~Mercuri, and E.~Rroji.
\newblock Approximation of the variance gamma model with a finite mixture of
  normals.
\newblock {\em Statistics \& Probability Letters}, 82(2):217 -- 224, 2012.

\bibitem{Granzer2013Thesis}
Granzer M.
\newblock {\em Estimation of COGARCH Models with implementation in R.}, 2013.
\newblock Master Thesis.

\bibitem{Madan1990}
D.~B. Madan and E.~Seneta.
\newblock The variance gamma (v.g.) model for share market returns.
\newblock {\em The Journal of Business}, 63(4):511--24, 1990.

\bibitem{maller2008}
R.~A. Maller, G.~M{\"u}ller, and A.~Szimayer.
\newblock Garch modelling in continuous time for irregularly spaced time series
  data.
\newblock {\em Bernoulli}, 14(2):519--542, 05 2008.

\bibitem{muller2010mcmc}
G.~M{\"u}ller.
\newblock Mcmc estimation of the cogarch (1, 1) model.
\newblock {\em Journal of Financial Econometrics}, 8(4):481--510, 2010.

\bibitem{newey1994large}
Whitney~K Newey and Daniel McFadden.
\newblock Large sample estimation and hypothesis testing.
\newblock {\em Handbook of econometrics}, 4:2111--2245, 1994.

\bibitem{protter1990stochastic}
P.~Protter.
\newblock {\em Stochastic integration and differential equations}.
\newblock Springer, 1990.

\bibitem{yuimaPack}
YUIMA~Project Team.
\newblock {\em yuima: The YUIMA Project package (stable version)}, 2013.
\newblock R package version 1.0.2.

\bibitem{Tomasson2011}
H.~Tomasson.
\newblock Some computational aspects of gaussian carma modelling.
\newblock {\em Statistics and Computing}, pages 1--13, 2013.

\bibitem{Tsai2005}
H.~Tsai and K.~S. Chan.
\newblock A note on non-negative continuous time processes.
\newblock {\em Journal of the Royal Statistical Society: Series B (Statistical
  Methodology)}, 67(4):589--597, 2005.

\end{thebibliography}



\end{document}